\documentclass[11pt,showpacs,showkeys,aps,pra]{revtex4}
\usepackage[utf8x]{inputenc}
\usepackage{ucs}

\usepackage{amsmath,amssymb}
\usepackage{latexsym}
\usepackage{amsfonts}
\usepackage{dcolumn}
\usepackage{bm}
\usepackage[usenames]{color}
\usepackage{multirow}
\usepackage{graphicx}
\usepackage{hyperref}
\usepackage{subfig}
\usepackage{booktabs}
\usepackage{xcolor}
\usepackage[normalem]{ulem}
\usepackage{float} 
\usepackage{wrapfig} 
\usepackage{upgreek} 
\usepackage{cancel} 
\usepackage{mathdots} 
\usepackage{mathrsfs} 
\usepackage{booktabs}

\newcommand{\doubt}{\marginpar{\fbox{\Large ?}}}

\begin{document}

\title{Heisenberg-like uncertainty measures for $D$-dimensional hydrogenic systems at large D}


\author{I.V. Toranzo}
\email[]{ivtoranzo@ugr.es}
\affiliation{Departamento de F\'{\i}sica At\'{o}mica, Molecular y Nuclear, Universidad de Granada, Granada 18071, Spain, and\\
Instituto Carlos I de F\'{\i}sica Te\'orica y Computacional, Universidad de Granada, Granada 18071, Spain}

\author{A. Martínez-Finkelshtein}
\email[]{andrei@ual.es}
\affiliation{Departamento de Matemáticas, Universidad de Almería, Almería 04120, Spain}
\affiliation{Instituto Carlos I de F\'{\i}sica Te\'orica y Computacional, Universidad de Granada, Granada 18071, Spain}

\author{J.S. Dehesa}
\email[]{dehesa@ugr.es}
\affiliation{Departamento de F\'{\i}sica At\'{o}mica, Molecular y Nuclear, Universidad de Granada, Granada 18071, Spain}
\affiliation{Instituto Carlos I de F\'{\i}sica Te\'orica y Computacional, Universidad de Granada, Granada 18071, Spain}

\begin{abstract}
The radial expectation values of the probability density of a quantum system in position and momentum spaces allow one to describe numerous physical quantities of the system as well as to find generalized Heisenberg-like uncertainty relations and to bound entropic uncertainty measures. It is known that the position and momentum expectation values of the  main prototype of the $D$-dimensional Coulomb systems, the $D$-dimensional hydrogenic system, can be expressed in terms of some generalized hypergeometric functions of the type $_{p+1}F_p(z)$ evaluated at unity with $p=2$ and $p=3$, respectively. In this work we determine the position and momentum expectation values in the limit of large $D$ for all hydrogenic states from ground to very excited (Rydberg) ones in terms of the spatial dimensionality and the hyperquantum numbers of the state under consideration. This is done by means of two different approaches to calculate the leading term of the special functions $_{3}F_2\left(1\right)$ and $_{5}F_4\left(1\right)$ involved in the large $D$ limit of the position and momentum quantities. Then, these quantities are used to obtain the generalized Heisenberg-like and logarithmic uncertainty relations, and some upper and lower bounds to the entropic uncertainty measures (Shannon, Rényi, Tsallis) of the $D$-dimensional hydrogenic system. 

\end{abstract}



\maketitle

\section{Introduction}

The study of the behavior of the physical properties of a $D$-dimensional system in terms of $D$ has a rich history in quantum mechanics and quantum field theory \cite{witten,yaffe1,herschbach,tsipis,chatterjee,avery,svidzinsky,dong} and, more recently, in quantum information \cite{bellomo, krenn,crann}. It has been observed that the physical phenomena depend on the dimension in a delicate way. For instance, the Huyghens principle of the wave propagation holds only when the spatial dimension is odd, while it is observed anomalous dispersion for any other real value of $D$ \cite{bender1,bender2,bender3}. Moreover, it is often possible to approximate the solution of difficult physical problems at the standard dimensionality ($D=3$) by means of a Taylor-series development of similar systems with a non-standard dimensionality in powers of $1/D$. This was motivated by the observation  that physics is much simpler when $D\rightarrow \infty$. This is true for a large variety of quantum systems from the single-particle systems moving in a $D$-dimensional central potential to more complex systems and phenomena (e.g. Casimir effects, random walks, and certain quantum field models containing $SU(D)$ gauge fields \cite{yaffe1,beldjenna,bender3}) as well as for quantum state tomography and some quantum codes and channnels \cite{crann}.\\[2mm]
Most relevant for our purposes is the development of the dimensional scaling method \cite{herschbach,tsipis} in the theory of many-electron systems, which offer novel, powerful and useful computational strategies for treating non-separable problems involving strong dynamical interactions \cite{yaffe2,chatterjee,avery,herschbach_87,herschbach96}. This method tipically starts with the generalization of the standard (three-dimensional) problem to a $D$-dimensional one and the introduction of a suitably scaled space to remove the major, generic $D$-dependence of the quantity under consideration; then, the evaluation of the scaled quantity at a large $D$ value, such as the limit $D\rightarrow \infty$, is performed in a relatively ``easy" way and finally one obtains an approximation for the standard value by relating it to this large-$D$-value by means of some interpolation or extrapolation procedure \cite{herschbach,herschbach_2000}. In the \textit{pseudoclassical} limit $D\rightarrow \infty$ of a many-electron system, which is tantamount to $h \rightarrow 0$ and/or $m_e \rightarrow \infty$ in the kinetic energy, the electrons assume fixed positions relative to the nuclei and each other in the $D$-scaled space \cite{herschbach86}. The large $D$ electronic geometry and energy correspond to the minimum of an exactly known effective potential and can be determined from classical electrostatics for any atom or molecule. For $D$ finite but very large, the electrons are confined to harmonic oscillations about the fixed positions attained in the $D\rightarrow\infty$ limit. Briefly, the large $D$ limit of numerous physical properties of almost all atoms with up to $100$ electrons and many diatomic molecules have been numerically evaluated, obtaining values comparable to or better than single-zeta Hartree-Fock calculations \cite{herschbach,herschbach96,tsipis}.\\[2mm]
Despite all these efforts the large $D$ limit of the main prototype of the Coulomb systems, the $D$-dimensional hydrogenic system, poses some open problems which can be solved analytically. It is known that the introduction of a $D$-dependent length scale converts the large $D$ limit of the associated Schr\"odinger equation into Bohr’s model \cite{svidzinsky}. On the other hand, we should keep in mind that the $D$-dimensional hydrogenic system (i.e., a negatively-charged particle moving in a space of $D$ dimensions around a positively charged core which electromagnetically binds it in its orbit) includes a wide variety of quantum systems, such as hydrogenic atoms and ions, exotic atoms, antimatter atoms, excitons, qubits,\dots

Moreover, the electronic distribution of the $D$-dimensional hydrogenic system is known (see next section) to have such a form that one can analytically determine its moments around the origin (radial expectation values) in both position \cite{pasternack,ray,andrae,drake,tarasov,guerrero11} and momentum \cite{hey2,assche} spaces as well as its entropic and complexity measures \cite{dehesa10,dehesa12}. These quantities describe and/or are closely related to various fundamental and/or experimentally accesible quantities (e.g., the diamagnetic susceptibility, the kinetic energy, the height peak of the Compton profile, the total electron-electron repulsion energy,\dots) and they characterize some position-momentum uncertainty-like relationships of Heisenberg \cite{zozor11,toranzo16} and entropic \cite{guerrero11} types (see also \cite{dehesa10} and references therein).\\[2mm]
Recently the radial expectation values of the $D$-dimensional hydrogenic states lying at the highest extreme region of the energy spectrum for a fixed $D$ were determined in both position and momentum spaces in terms of $D$ and the state's hyperquantum principal and orbital quantum numbers \cite{aptekarev}. In this work we analytically determine these position and momentum quantities for all quantum hydrogenic states in the large dimensionality limit. First, in Section II the known physical solutions of the Schr\"odinger equation of the $D$-dimensional hydrogenic system are given in the two conjugated spaces, as well as the associated position and momentum probability densities and their corresponding radial expectation values in terms of the space dimensionality and of the hyperquantum numbers which characterize the system's states. We will see that the position and momentum expectation values are expressed in terms of some generalized hypergeometric functions \cite{nist} of the type $_{p+1}F_p\left(a_{1},\ldots,a_{p+1}; b_{1},\ldots,b_{p} ;z\right)$ evaluated at $z=1$, with $p=2$ and $p=3$, respectivley. Then, in Sections III and IV the position and momentum expectation values of the system are evaluated in the large dimensionality limit for the ground and excited states, respectively. Two different asymptotic approaches are developed to calculate the dominant term of the special functions $_{3}F_2\left(a_{1},a_{2},a_{3}; b_{1},b_{2} ;1\right)$ and $_{5}F_4\left(a_{1},a_{2},a_{3},a_{4},a_{5}; b_{1},b_{2},b_{3},b_{4} ;1\right)$ involved in the large $D$ limit of the position and momentum quantities under consideration. In Section V the position and momentum expectation values of the Rydberg (i.e., large $n$) hydrogenic states are calculated in the large $D$ limit. In Section VI, we give the uncertainty relations of Heisenberg and logarithmic types for all the stationary states of a $D$-dimensional hydrogenic system at the large-$D$ limit, and we show that they fulfill and saturate the general inequality-type uncertainty relations of all quantum systems. In Section VII, the position and momentum expectation values are shown to bound the Shannon and Rényi entropic uncertainty measures from above, and the Tsallis' uncertainty measure from below. Finally, some conclusions and open problems are given.

\section{The $D$-dimensional hydrogenic densities in position and momentum spaces}

In this section we briefly describe the wavefunctions of the ground and excited states of the $D$-dimensional hydrogenic system and the associated electronic distribution densities in the two conjugated position and momentum spaces, as well as the exact, compact values of their radial and logarithmic expectation values.\\
\subsection{Position space} 
The time-independent Schr\"{o}dinger equation of a $D$-dimensional ($D \geqslant 1$) hydrogenic system (i.e., an electron moving under the action of the $D$-dimensional Coulomb potential $\displaystyle{V(\vec{r})=-\frac{Z}{r}}$) is given by
\begin{equation}\label{eqI_cap1:ec_schrodinger}
\left( -\frac{1}{2} \vec{\nabla}^{2}_{D} - \frac{Z}{r}
\right) \Psi \left( \vec{r} \right) = E \Psi \left(\vec{r} \right),
\end{equation}
where $\vec{\nabla}_{D}$ denotes the $D$-dimensional gradient operator, $Z$ is the nuclear charge, and the  electronic position vector  is  given in hyperspherical units  as $\vec{r} = (r,\theta_1,\theta_2,\ldots,\theta_{D-1})      \equiv
(r,\Omega_{D-1})$, $\Omega_{D-1}\in S^{D-1}$, where $r \equiv |\vec{r}| \in [0  \: ;  \: +\infty)$  and 
$\theta_i \in [0 \: ; \: \pi), i < D-1$, $\theta_{D-1} \equiv \phi \in [0 \: ; \: 2
\pi)$. It is assumed that the nucleus is located at the origin and, by  convention, $\theta_D =  0$ and the  empty product is the  unity. Atomic units are used throughout the paper.\\
It is known \cite{nieto,yanez,avery,dehesa10} that the energies belonging to the discrete spectrum are given by  
\begin{equation} \label{eqI_cap1:energia}
E= -\frac{Z^2}{2\eta^2},\hspace{0.5cm} \hspace{0.5cm} \eta=n+\frac{D-3}{2}; \hspace{5mm} n=1,2,3,...,
\end{equation}
and the associated eigenfunction can be expressed as
\begin{equation}\label{eqI_cap1:FunOnda_P}
\Psi_{\eta,l, \left\lbrace \mu \right\rbrace }(\vec{r})=\mathcal{R}_{\eta,l}(r)\,\, {\cal{Y}}_{l,\{\mu\}}(\Omega_{D-1}),
\end{equation}
where $(l,\left\lbrace \mu \right\rbrace)\equiv(l\equiv\mu_1,\mu_2,...,\mu_{D-1})$ denote the hyperquantum numbers associated to the angular variables $\Omega_{d-1}\equiv (\theta_1, \theta_2,...,\theta_{D-1})$, which may take all values consistent with the inequalities $l\equiv\mu_1\geq\mu_2\geq...\geq \left|\mu_{D-1} \right| \equiv \left|m\right|\geq 0$. The radial eigenfunction is given by
\begin{align}\label{eqI_cap1:Rnl}
\mathcal{R}_{n,l}(r)&=K_{n,l}\left(\frac{r}{\lambda}\right)^{l}e^{-\frac{r}{2\lambda}} \mathcal{L}_{n-l-1}^{(2l+D-2)}\left(\frac{r}{\lambda}\right)\\ \nonumber
&=K_{n,l}\left[\frac{\omega_{2L+1}(\tilde{r})}{\tilde{r}^{D-2}}\right]^{1/2}{\cal{L}}_{\eta-L-1}^{(2L+1)}(\tilde{r})\\ \nonumber
&= \left( \frac{\lambda^{-D}}{2 \eta}\right)^{1/2}   \left[\frac{\omega_{2L+1}(\tilde{r})}{\tilde{r}^{D-2}}\right]^{1/2}{\widehat{\cal{L}}}_{\eta-L-1}^{(2L+1)}(\tilde{r}),
\end{align}
where the ``grand orbital angular momentum quantum number'' $L$ and the adimensional parameter $\tilde{r}$ are
\begin{align} \label{eqI_cap1:Lyr}
L&=l+\frac{D-3}{2}, \hspace{0.5cm} l=0, 1, 2, \ldots \\ \label{rtilde}
\tilde{r}&=\frac{r}{\lambda},\hspace{0.5cm} 
\hspace{0.5cm}\lambda=\frac{\eta}{2Z},
\end{align}
and
$\omega_{\beta}(x) =x^{\beta}e^{-x}, \, \beta =2l+D-2,$
  is the weight function of the Laguerre polynomials with parameter $\beta$. The symbols $\mathcal{L}_{n}^{(\beta)}(x)$ and $\widehat{\mathcal{L}}_{n}^{(\beta)}(x)$ denote the orthogonal and orthonormal, respectively, Laguerre polynomials with respect to the weight $\omega_\beta(x)=x^{\beta} e^{-x}$ on the interval $\left[0,\infty \right) $, so that
\begin{equation}\label{eqI_cap1:laguerre_orto_ortogo}
{\widehat{\mathcal{L}}}^{(\beta)}_{m}(x)=  \left( \frac{m!}{\Gamma(m+\beta+1)}\right)^{1/2}  {\mathcal{L}}^{(\beta)}_{m}(x),
\end{equation}
and finally
\begin{equation}
K_{n,l}  = 
 \lambda^{-\frac{D}{2}}\left\{\frac{(\eta-L-1)!}{2\eta(\eta+L)!}\right\}^{\frac{1}{2}}=\left\{\left(\frac{2Z}{n+\frac{D-3}{2}}\right)^{D}\frac{(n-l-1)!}{2\left(n+\frac{D-3}{2}\right)(n+l+D-3)!}  \right\}^{\frac{1}{2}} 
\label{eq:4}
\end{equation}
is the normalization constant which ensures that $\int \left| \Psi_{\eta,l, \left\lbrace \mu \right\rbrace }(\vec{r}) \right|^2 d\vec{r} =1$. 
The angular eigenfunctions are the hyperspherical harmonics, $\mathcal{Y}_{l,\{\mu\}}(\Omega_{D-1})$, defined as
\begin{equation}
\label{eq:hyperspherarm}
\mathcal{Y}_{l,\{\mu\}}(\Omega_{D-1}) = \mathcal{N}_{l,\{\mu\}}e^{im\phi}\nonumber\times \prod_{j=1}^{D-2}\mathcal{C}^{(\alpha_{j}+\mu_{j+1})}_{\mu_{j}-\mu_{j+1}}(\cos\theta_{j})(\sin\theta_{j})^{\mu_{j+1}}
\end{equation}
with the normalization constant
\begin{equation}
\label{eq:normhypersphar}
\mathcal{N}_{l,\{\mu\}}^{2} = \frac{1}{2\pi}\times\nonumber\\
\prod_{j=1}^{D-2} \frac{(\alpha_{j}+\mu_{j})(\mu_{j}-\mu_{j+1})![\Gamma(\alpha_{j}+\mu_{j+1})]^{2}}{\pi \, 2^{1-2\alpha_{j}-2\mu_{j+1}}\Gamma(2\alpha_{j}+\mu_{j}+\mu_{j+1})},\nonumber\\
\end{equation}
where the symbol $\mathcal{C}^{(\lambda)}_{n}(t)$ denotes the Gegenbauer polynomial of degree $n$ and parameter $\lambda$, orthogonal on [-1,1] with respect to the weight function $w_{\nu}(t) = (1-t^2)^{\nu-1/2}$.\\[2mm]

The quantum probability density of a $D$-dimensional hydrogenic stationary state $(n,l,\{\mu\})$ is the square of the absolute value of the position eigenfunction, 
\begin{equation}
\label{eq:denspos}
\rho_{n,l,\{\mu\}}(\vec{r}) = \rho_{n,l}(\tilde{r})\,\, |\mathcal{Y}_{l,\{\mu\}}(\Omega_{D-1})|^{2},
\end{equation}
where the radial part of the density is the univariate function
\begin{eqnarray}
\label{eq:radensity}
\rho_{n,l}(\tilde{r}) &=& [\mathcal{R}_{n,l}(r)]^2 = \frac{\lambda^{-D}}{2 \eta} \,\, \frac{\omega_{2L+1}(\tilde{r})}{\tilde{r}^{D-2}}\,\,[{\widehat{\mathcal{L}}}_{\eta-L-1}^{(2L+1)}(\tilde{r})]^{2}.
\end{eqnarray}
The moments (centered at the origin) of this density function are the radial expectation values in the position space, and can be expressed in the following compact form \cite{andrae,drake,tarasov,dehesa10}:
\begin{eqnarray}
\label{eq:radexpec1}
\langle r^{\alpha} \rangle &=& \int r^{\alpha}\rho_{n,l,\{\mu \}}(\vec{r})\, d\vec{r} = \int_{0}^{\infty} r^{\alpha+D-1}\rho_{n,l}(\tilde{r})\, dr \nonumber  \\
\label{eq:radexpec1bis}
&=&
 \frac{1}{2\eta}\left(\frac{\eta}{2Z}\right)^{\alpha}\int_{0}^{\infty} w_{2l+D-2}(\tilde{r})[\mathcal{\widehat{L}}^{(2l+D-2)}_{n-l-1}(\tilde{r})]^{2}\, \tilde{r}^{\alpha+1}\, d\tilde{r} \\
 \label{eq:radexpec2}
&=&  \frac{\eta^{\alpha-1}}{2^{\alpha+1}Z^{\alpha}}\frac{\Gamma(2L+\alpha+3)}{\Gamma(2L+2)} \nonumber\\
& & \times \,\, _3F_2(-\eta+L+1,-\alpha-1,\alpha+2;2L+2,1;1),
\end{eqnarray}
which holds for $\alpha > -D-2l$.
Notice that these quantities are given in terms of $\eta,L$ and the nuclear charge $Z$. In particular, we have $\langle r^{0} \rangle = 1$, as well as the following values for the first few negative and positive expectation values:
\begin{eqnarray}
\label{eq:rvar}
\langle r^{-1} \rangle &=& \frac{Z}{\eta^{2}},\quad 
\langle r\rangle = \frac{1}{2Z}[3\eta^{2}-L(L+1)] , \quad
\langle  r^{2} \rangle = \frac{\eta^{2}}{2Z^{2}}[5\eta^{2}+1-3L(L+1)],\quad
\langle r^{-2} \rangle = \frac{Z^{2}}{\eta^{3}}\frac{1}{L+\frac{1}{2}} \nonumber \\
\langle r^{-3} \rangle &=& \frac{Z^{3}}{\eta^{3}L(L+\frac{1}{2})(L+1)}, \quad
\langle r^{-4} \rangle = Z^{4}\frac{3\eta^{2}-L(L+1)}{2\eta^{5}(L-\frac{1}{2})L(L+\frac{1}{2})(L+1)(L+\frac{3}{2})}. 
\end{eqnarray}
Furthermore, the radial logarithmic values \cite{dehesa10} are given by 
\begin{eqnarray}
\label{eq:logexpec}
\langle \log r \rangle &=& \int (\log r) \rho_{n,l,\{\mu \}}(\vec{r})\, d\vec{r} \nonumber \\
&=&  \log\left(n+\frac{D-3}{2}\right)+\frac{2n-2l-1}{2n+D-3}+\psi(n+l+D-2)-\log(2Z),
\end{eqnarray}
where $\psi(x)\equiv \Gamma'(x)/\Gamma(x)$ is the digamma function \cite{nist}.\\

Some relevant particular cases are:
\begin{itemize}
\item For the ground state ($n=1, l=0$) we obtain the expressions
\begin{eqnarray}
\label{eq:radpart}
\langle r^{\alpha} \rangle &=& \left(\frac{D-1}{4Z}\right)^{\alpha}\frac{\Gamma (D+\alpha )}{\Gamma (D)}; \quad \alpha > -D  \\
\langle \log r \rangle &=&  \psi(D)+ \log(D-1) - 2\log 2 - \log Z
\end{eqnarray}
for the radial, both conventional and logarithmic, expectation values in position space, respectively.
\item For Rydberg states ($n\gg 1$), the radial expectation values have been recently shown \cite{aptekarev} to be given as
\begin{equation}
\label{eq:rydr}
\langle r^{\alpha} \rangle = \left(\frac{\eta^{2}}{Z}\right)^{\alpha} \,\,\frac{2^{\alpha+1}\,\,\Gamma(\alpha+\frac{3}{2})}{\sqrt{\pi}\,\,\Gamma(\alpha+2)}\left(1+o(1)\right), \quad n\rightarrow \infty
\end{equation}
with $(\alpha,l,D)$ uniformly bounded and $\alpha >-3/2$.
\end{itemize}

\subsection{Momentum space}

In momentum space we can work out similarly the corresponding Schr\"odinger equation of our system to find \cite{aquilanti,yanez,szmytkowski} the following expression for the momentum wavefunction of the $D$-dimensional hydrogenic stationary state $(n,l,\{\mu\})$:
\begin{equation}
\label{eq:momwvf}
\tilde{\Psi}(\vec{p}) = \mathcal{M}_{n,l}(p)\mathcal{Y}_{l,\{\mu \}}(\Omega_{D-1}),
\end{equation}
where the radial momentum wavefunction is 
\begin{equation}
\label{eq:radmomwvf}
\mathcal{M}_{n,l}(p) = K_{n,l}\, \frac{(\eta \tilde{p})^{l}}{(1+\eta^{2}\tilde{p}^{2})^{L+2}}\, \mathcal{C}_{\eta-L-1}^{(L+1)}\left( \frac{1-\eta^{2}\tilde{p}^{2}}{1+\eta^{2}\tilde{p}^{2}} \right),
\end{equation}
with $\tilde{p} =p/Z$, and the normalization constant
\begin{equation}
\label{eq:normacons}
K_{n,l} = Z^{-\frac{D}{2}}2^{2L+3}\left[\frac{(\eta-L-1)!}{2\pi(\eta+L)!} \right]^{\frac{1}{2}}\Gamma(L+1)\eta^{\frac{D+1}{2}}.
\end{equation}
Then, the momentum probability density is 
\begin{eqnarray}
\label{eq:momdens}
\gamma_{n,l,\{\mu \}}(\vec{p}) &=& |\tilde{\Psi}_{n,l,\{\mu \}}(\vec{p})|^{2} =  \mathcal{M}^{2}_{n,l}(p) |\mathcal{Y}_{l,\{\mu \}}(\Omega_{D-1})|^{2}\nonumber \\
&=& K_{n,l}^{2}\frac{(\eta \tilde{p})^{2l}}{(1+\eta^{2}\tilde{p}^{2})^{2L+4}}\left[\mathcal{C}_{\eta-L-1}^{(L+1)}\left( \frac{1-\eta^{2}\tilde{p}^{2}}{1+\eta^{2}\tilde{p}^{2}} \right)\right]^{2}|\mathcal{Y}_{l,\{\mu \}}(\Omega_{D-1})|^{2}.
\end{eqnarray}
The moments centered at the origin of this density function are the radial expectation values in the momentum space, which can be expressed in the following compact form \cite{hey2,assche}:
\begin{eqnarray}
\label{eq:momexpec}
\langle p^{\alpha} \rangle &=& \int p^{\alpha}\gamma_{n,l,\{\mu \}}(\vec{p})\, d\vec{p} = \int_{0}^{\infty} p^{\alpha+D-1}\mathcal{M}^{2}_{n,l}(p)\, dp  \nonumber \\
&=& \left(\frac{Z}{\eta}\right)^{\alpha} \int_{-1}^{1} w_{\nu}(t)[\widehat{\mathcal{C}}_{k}^{(\nu)}(t)]^{2}(1-t)^{\frac{\alpha}{2}}(1+t)^{1-\frac{\alpha}{2}}\,dt\nonumber  \\
\label{eq:momint}
&=& \left(\frac{Z}{\eta}\right)^{\alpha}\frac{2^{2\nu-1}k!(k+\nu)}{\pi \Gamma(k+2\nu)} [\Gamma(\nu)]^{2} \int_{-1}^{1} w_{\nu}(t)[\mathcal{C}_{k}^{(\nu)}(t)]^{2}(1-t)^{\frac{\alpha}{2}}(1+t)^{1-\frac{\alpha}{2}}\,dt \\
& = & \frac{2^{1-2\nu}Z^{\alpha}\sqrt{\pi}}{k!\,\eta^{\alpha}}\frac{(k+\nu)\Gamma(k+2\nu)\Gamma(\nu+\frac{\alpha+1}{2})\Gamma(\nu+\frac{3-\alpha}{2})}{\Gamma^{2}(\nu+\frac{1}{2})\Gamma(\nu+1)\Gamma(\nu+\frac{3}{2})}\nonumber \\
& & \times \,\, _5F_4(-k,k+2\nu,\nu,\nu+\frac{\alpha+1}{2},\nu+\frac{3-\alpha}{2};2\nu,\nu+\frac{1}{2},\nu+1,\nu+\frac{3}{2};1),
\end{eqnarray} 
which holds for $\alpha \in (-D-2l, D+2l+2)$. Here the notations $k = \eta + L +1 = n-l-1$ and $\nu = L+1 = l + (D-1)/2$ have been used. Moreover, the symbol $\widehat{\mathcal{C}}^{(\lambda)}_{m}(t)$ denotes the orthonormal Gegenbauer polynomials so that
\begin{equation}
\label{eq:gegen}
\widehat{\mathcal{C}}^{(\lambda)}_{m}(t) = \frac{m!(m+\lambda)[\Gamma(\lambda)]^{2}}{\pi\, 2^{1-2\lambda} \Gamma(m+2\lambda)} \,\mathcal{C}^{(\lambda)}_{m}(t) .
\end{equation}
 Observe again that  the momentum expectation values $\langle p^{\alpha} \rangle$ are given in terms of $\eta,L$ and the nuclear charge $Z$; or, equivalently, in terms of $n,l,D$ and $Z$. In particular, we have $\langle p^{0} \rangle = 1$, as well as the following expectation values with negative and positive even powers:
\begin{eqnarray}
\label{eq:pvar}
\langle p^{-2} \rangle &=& \frac{Z^{-2}}{\eta^{-2}}\frac{8\eta -3(2L+1)}{2L+1}, \quad 
\langle p^{2}\rangle = \frac{Z^{2}}{\eta^{2}}, 	\quad
\langle  p^{4} \rangle = \frac{Z^{4}}{\eta^{4}}\frac{8\eta -3(2L+1)}{2L+1}.\nonumber\\
\langle  p^{6} \rangle &=& \frac{Z^{6}}{\eta^{6}}\frac{(4k+2\nu+1)(16k^2+40\nu k-4k+ 4 \nu^2 + 16 \nu+ 15)}{(2L+3)(2L+1)(2L-1)}.
\end{eqnarray}
Moreover,  
\begin{equation}
\langle  p^{-\beta} \rangle	= \eta^{2\beta+2} \langle  p^{\beta+2} \rangle, \quad  \beta = 0,1,2,...
\end{equation}
Note that the expectation values with odd integer powers are not explicitly known, except possibly for the case $p=-1$, which has a somewhat complicated expression \cite{delbourgo}.\\

Furthermore, the logarithmic expectation values of the momentum density function are given by 
\begin{eqnarray} 
\label{eq:logmomexpec}
\langle \log p \rangle &=&
\int (\log p) \gamma_{n,l,\{\mu \}}(\vec{p})\, d\vec{p}\nonumber \\
&=&  -\log\left(n+\frac{D-3}{2}\right)+\frac{(2l+D-2)(2n+D-3)}{(2n+D-3)^{2}-1}-1+\log(Z).
\end{eqnarray}
Some relevant particular cases:
\begin{itemize}
	\item For the ground state one obtains 
\begin{eqnarray}
\label{eq:mompart}
\langle p^{\alpha} \rangle &=& \left(\frac{2Z}{D-1} \right)^{\alpha}\frac{2\Gamma(\frac{D-\alpha}{2}+1)\Gamma(\frac{D+\alpha}{2})}{D\Gamma^{2}\left(\frac{D}{2} \right)}, \quad -D<\alpha<D+2\\
\langle \log p \rangle &=& -\log(D-1) + \log 2 -\frac{1}{D} +\log Z
\end{eqnarray}
for the radial (conventional and logarithmic) values in momentum space, respectively.
\item For Rydberg states ($n\gg 1$), it has been recently shown \cite{aptekarev} that for $(l,D)$ uniformly bounded, the momentum expectation values satisfy
\begin{equation}
\label{eq:rydp1}
\langle p^{\alpha} \rangle \simeq \left(\frac{Z}{\eta}\right)^{\alpha} \left\{ \begin{array}{cc}
\frac{\alpha-1}{\sin(\pi(\alpha-1)/2)}, & -1<\alpha <3, \quad \alpha \neq 1, \\
2/\pi, & \alpha =1
\end{array}\right.
\end{equation}
understanding by $\simeq$  that the ratio of the left and right hand sides tends to $1$ as $n\to \infty$. 
Moreover, for Rydberg states such that both $n$ and $l$ tend to infinity with the condition $n-l =\text{constant}$, the radial momentum expectation values are given by
\begin{equation}
\label{eq:rydp2}
\langle p^{\alpha}\rangle \simeq \left(\frac{Z}{\eta}\right)^{\alpha} \frac{1}{2\pi}\int_{-1}^{1}\frac{(2-\sqrt{3}t)^{\frac{\alpha}{2}}(2+\sqrt{3}t)^{1-\frac{\alpha}{2}}}{\sqrt{1-t^{2}}}\, dt,
\end{equation} 
provided that $D$ is bounded.
\end{itemize}

\section{Position expectation values of large-$D$ hydrogenic systems}

In this section we calculate the position radial and logarithmic expectation values for an arbitrary (but fixed) state $(n,l,\{\mu \})$ of $D$-dimensional hydrogenic systems when $D\rightarrow \infty$. Let us first start with the radial expectation values $\langle r^{\alpha}\rangle$. We claim that these quantities have the following asymptotic expression:
\begin{eqnarray}
\label{eq:averk1}
\langle r^{\alpha} \rangle &= & \left(\frac{D^2}{4 Z}\right)^{\alpha }\left(1+\frac{(\alpha+1)(\alpha+4l-2)}{2D} \right)\left(1+\frac{(\alpha +1) (\alpha +2) (n-l-1)}{D+2 l-1}\right)  \left(1+\mathcal{O}\left(\frac{1}{D^{2}} \right)\right)\nonumber \\
\end{eqnarray}
as $D\rightarrow \infty$, which holds for $\alpha > -D-2l$. Notice that in such limit one has that $\left(\frac{D^2}{4 Z}\right)^{-\alpha}\langle r^{\alpha} \rangle \rightarrow 1$. Thus, our $D$-dimensional hydrogenic system has a characteristic length, $r_{char} = \frac{D^2}{4 Z}$, which corresponds to the localization of the maximum of the ground-state probability density. Moreover, it is the radial distance at which the effective potential attains a minimum as $D\rightarrow \infty$. Therefore, the electron of the $D$-dimensional hydrogenic system behaves as it is moving in a circular orbit with radius $r_{char}$ and angular momentum $D/2$, experimenting quantum fluctuations from this orbit vanishing as $D^{-1/2}$, as it was previously noted by \cite{ray}, since
$$\frac{\Delta r}{\langle r \rangle} = \frac{(\langle r^{2} \rangle - \langle r \rangle^{2})^{1/2}}{\langle r \rangle} = \frac{1}{\sqrt D}. $$
For illustrative purposes, we show the rate of convergence of these large-$D$ values to the exact ones in Table \ref{tab:PPer}; see Appendix \ref{table:app}.

Let us now prove the main result (\ref{eq:averk1}). We start from the general expression (\ref{eq:radexpec1})-(\ref{eq:radexpec2}) for the radial expectation value of the $D$-hydrogenic state $(n,l,\{\mu \})$, which is given in terms of the generalized hypergeometric function $_{3}F_2$ evaluated at $1$, and then we use the following asymptotic expression of $_{p+1}F_p$  for large parameters (see  \cite{knotterus},  \cite[Eq. (7.3)]{luke} or \cite[Eq. (16.11.10)]{nist}):
\begin{eqnarray}
\label{eq:ashyp}
_{p+1}F_p\left(
				a_{1}+r,\ldots,a_{k-1}+r,a_{k},\ldots,a_{p+1};  
				b_{1}+r,\ldots,b_{k}+r,b_{k+1},\ldots,b_{p} ;z\right)	 &=& \nonumber \\
			=	\sum_{j=0}^{m-1}\frac{(a_{1}+r)_{j}\cdots(a_{k-1}+r)_{j}(a_{k})_{j}\cdots(a_{p+1})_{j}}{(b_{1}+r)_{j}\cdots(b_{k}+r)_{j}(b_{k+1})_{j}\cdots(b_{p})_{j}}\frac{z^{j}}{j!} + \mathcal O\left(\frac{1}{r^{m}}\right) & &
\end{eqnarray}
as $r\rightarrow +\infty$, where $z$ is fixed, \doubt $|ph(1-z)|<\pi$, $m\in \mathbb{Z}^{+}$, and $k$ can take any integer value from $1$ to $p$. We have also used the Pochhammer symbol $(a)_j = \Gamma(a+j)/\Gamma(a)$. With $p=2$ and $k=1$ we obtain the following asymptotics for the $_{3}F_2$ hypergeometric function of our interest:
\begin{eqnarray}
\label{eq:3F2asymp0}
_{3}F_2\left(
				a_{1},a_{2},a_{3};
				b_{1}+r,b_{2} ;z\right)	 &=& 
				\sum_{j=0}^{m-1}\frac{(a_{1})_{j}(a_{2})_{j}(a_{3})_{j}}{(b_{1}+r)_{j}(b_{2})_{j}}\frac{z^{j}}{j!} + \mathcal O\left(\frac{1}{r^{m}}\right)
\end{eqnarray}
Applying this expression in (\ref{eq:radexpec2}) with $z=1$ and $r = D$, one has
\begin{equation}
\label{eq:3F2asymp1}
_{3}F_2\left(
				-n+l+1,-\alpha-1,\alpha+2; 
			2l-1+D,1;1\right)	=	\sum_{j=0}^{m-1}\frac{(-n+l+1)_{j}(-\alpha-1)_{j}(\alpha+2)_{j}}{(2l-1+D)_{j}(1)_{j}}\frac{1}{j!} + \mathcal O\left(\frac{1}{D^{m}}\right), 
\end{equation}
which, when $D\rightarrow +\infty$, yields for $m=2$ the asymptotics
\begin{equation}
\label{eq:3F2asymp2}
_{3}F_2\left(
				-n+l+1,-\alpha-1,\alpha+2; 
			2l-1+D,1;1\right) = 	1+ \frac{(\alpha +1) (\alpha +2) (n-l-1)}{2 l-1+D} + \mathcal O\left(\frac{1}{D^{2}}\right).
\end{equation}
Now, by taking into account Eq. (\ref{eq:radexpec2}) together with (\ref{eq:3F2asymp2}) and the following asymptotics of the ratio (see e.g.,\cite[Eq. (5.11.12)]{nist})
\begin{equation}
\label{eq:gammapprox}
\frac{\Gamma (D+2 l+\alpha )}{\Gamma (D+2 l-1)} = D^{1+\alpha}\left(1+\frac{(\alpha+1)(\alpha+4l-2)}{2D}+ \mathcal O\left( \frac{1}{D^{2}} \right)\right),
\end{equation}
we have 
\begin{eqnarray}
\label{eq:aver}
\langle r^{\alpha} \rangle &= & \left(\frac{D^2}{4 Z}\right)^{\alpha } \left(1+\frac{(\alpha+1)(\alpha+4l-2)}{2D} \right)\left(1+\frac{(\alpha +1) (\alpha +2) (n-l-1)}{D+2 l-1}\right) \left(1+\mathcal{O}\left(\frac{1}{D^{2}}\right)  \right)\nonumber \\
&= & \left(\frac{D^2}{4 Z}\right)^{\alpha } \left(1+\frac{(\alpha+1)(\alpha+4l-2)}{2D} \right)\left(1+\frac{(\alpha +1) (\alpha +2) (n-l-1)}{D}\right)\left(1+\mathcal{O}\left(\frac{1}{D^{2}}\right)  \right),\nonumber \\
\end{eqnarray}
which gives the expression (\ref{eq:averk1}).

Now, let us explore the behavior of the logarithmic expectation value (\ref{eq:logexpec}) of the $D$-dimensional hydrogenic system for large $D$. Taking into account that $\psi(z) = \log z -\frac{1}{2z} + \mathcal{O}\left(\frac{1}{z^{2}}\right)$ for $z \rightarrow  \infty$ (see e.g., \cite[Eq. (25.16.3)]{nist}) and $\log(a+bz) = \log(bz)+\frac{a}{bz} + \mathcal{O}\left(\frac{1}{z^{2}}\right)$ for $z\rightarrow \infty$, one has from (\ref{eq:logexpec}) that 
\begin{eqnarray}
\label{eq:avelogr}
\langle \log r \rangle &= & 2 \log D -\log (4 Z) + \frac{5 n-l-\frac{13}{2}}{D} + \mathcal{O}\left(\frac{1}{D} \right).
\end{eqnarray}
Finally, for circular states ($l=n-1$) one has that the position and logarithmic expectation values given by (\ref{eq:averk1}) and (\ref{eq:avelogr}), respectively, reduce to
\begin{equation}
\label{eq:averacirc}
\langle r^{\alpha} \rangle_{cs} = \left(\frac{D^2}{4Z}\right)^{\alpha } \left[1+\frac{(\alpha +1) (4n+\alpha -6)}{2 D}\right]\left( 1+ \mathcal{O}\left(\frac{1}{D^{2}} \right)\right)
\end{equation}
and
\begin{equation}
\label{eq:avelogrcirc}
\langle \log r \rangle_{cs} = \frac{4n-\frac{11}{2}}{D} + 2 \log D -\log (4 Z)+ \mathcal{O}\left(\frac{1}{D} \right),
\end{equation}
respectively. Moreover, from these expressions we can easily obtain the position and logarithmic expectation values for the ground state ($n=1$) of the $D$-dimensional hydrogenic system at large $D$.

\section{Momentum expectation values of large-$D$ hydrogenic systems}

In this section we calculate the momentum radial and logarithmic expectation values for an arbitrary (but fixed) state $(n,l,\{\mu \})$ of $D$-dimensional hydrogenic systems when $D\rightarrow \infty$. First, let us consider the momentum expectation values $\langle p^{\alpha}\rangle$. We claim that these quantities have the following asymptotic expression:
\begin{eqnarray}
\label{eq:avepa}
\langle p^{\alpha} \rangle &=&\left(\frac{Z}{n+\frac{D-3}{2}} \right)^{\alpha}\left(1+\frac{\alpha(\alpha-2)(2n-2l-1)}{2D} + \mathcal{O}(D^{-2})\right) \nonumber \\
& = & \left(\frac{2Z}{D} \right)^{\alpha}\left(1+\frac{\alpha(\alpha-2)(2n-2l-1)}{2D} + \mathcal{O}(D^{-2})\right)
\end{eqnarray}
as $D\rightarrow \infty$, which holds for $\alpha \in (-D-2l, D+2l+2)$. Notice that in such limit one has that $\left(\frac{D^2}{4 Z}\right)^{-\alpha}\langle p^{\alpha} \rangle \rightarrow 1$. Thus, our $D$-dimensional hydrogenic system has a characteristic momentum, $p_{char} = \frac{D^2}{4 Z}$, which corresponds to the localization of the maximum of the ground-state probability density in momentum space. Moreover, it gives the velocity at which the electron of our system moves in the circular orbit defined in the previous section as $D\rightarrow \infty$. For illustrative purposes, we show the rate of convergence of these large-D values to the exact ones in Table \ref{tab:PPer}; see Appendix \ref{table:app}.\\

Let us now prove the main result (\ref{eq:avepa}). We start from  the general expression (\ref{eq:momexpec}). Then, using  the definition of the hypergeometric function and the duplication formula of the gamma function (see e.g., \cite{nist}), we can rewrite (\ref{eq:momexpec}) as
\begin{eqnarray}
\label{eq:1}
\langle p^{\alpha}\rangle \frac{\eta^{\alpha}}{Z^{\alpha}} &=& \frac{2}{k!}\frac{(k+\nu)\Gamma(k+2\nu)}{\Gamma(2\nu+1)}\frac{\Gamma(\nu+\frac{\alpha+1}{2})\Gamma(\nu+\frac{3-\alpha}{2})}{\Gamma(\nu+\frac{1}{2})\Gamma(\nu+\frac{3}{2})}\nonumber \\
&\times & \sum_{j=0}^{k}(-1)^{j}\binom{k}{j}\frac{(k+2\nu)_{j}(\nu)_{j}(\nu+\frac{\alpha+1}{2})_{j}(\nu+\frac{3-\alpha}{2})_{j}}{(2\nu)_{j}(\nu+1)_{j}(\nu+\frac{1}{2})_{j}(\nu+\frac{3}{2})_{j}}
\end{eqnarray}
We want to determine the asymptotics of this quantity in the $D\rightarrow \infty$ limit when $n$ and $l$ are fixed. Since $k=n-l-1$ and $\nu = l + (D-1)/2$, one realizes that we have to compute the asymptotics of (\ref{eq:1}) when $\nu\rightarrow \infty$ and $k$ is fixed. To begin with, we take into account the following identities for the Pochhammer symbols
\begin{eqnarray}
\frac{(2\nu+k)_{j}}{(2\nu)_{j}} =	\frac{(2\nu+j)_{j}}{(2\nu)_{j}}, \quad  \frac{(\nu)_{j}}{(\nu+1)_{j}} =	\frac{\nu}{\nu+j}
\end{eqnarray}
in Eq. (\ref{eq:1}), so that we can rewrite it as follows
\begin{eqnarray}
\label{eq:1bis}
\langle p^{\alpha}\rangle \frac{\eta^{\alpha}}{Z^{\alpha}} &=& \frac{2}{k!}\frac{(k+\nu)\Gamma(k+2\nu)}{\Gamma(2\nu+1)}\frac{\Gamma(\nu+\frac{\alpha+1}{2})\Gamma(\nu+\frac{3-\alpha}{2})}{\Gamma(\nu+\frac{1}{2})\Gamma(\nu+\frac{3}{2})}f_{k}(\nu)
\end{eqnarray}
with 
\begin{equation}
\label{eq:2}
f_{k}(\nu) = \frac{1}{(2\nu)_{k}}\sum_{j=0}^{k}(-1)^{j}\binom{k}{j}(2\nu+j)_{k}\,d_{j},
\end{equation}
where
\begin{eqnarray}
\label{eq:3}
d_{j}\equiv d_{j}(\nu) &=& \frac{\nu}{\nu+j}\frac{(\nu+\frac{\alpha+1}{2})_{j}(\nu+\frac{3-\alpha}{2})_{j}}{(\nu+\frac{1}{2})_{j}(\nu+\frac{3}{2})_{j}}.
\end{eqnarray}
In order to find the asymptotics of (\ref{eq:1bis}) we take into account that, as $\nu\rightarrow +\infty$,
\begin{equation}
\label{eq:11}
2\frac{(k+\nu)\Gamma(k+2\nu)}{\Gamma(2\nu+1)} = (2\nu)^{k}\left(1+\frac{k(k+3)}{4\nu}+o(1/\nu)\right).
\end{equation}
and
\begin{equation}
\label{eq:10}
\frac{\Gamma(\nu+\frac{\alpha+1}{2})\Gamma(\nu+\frac{3-\alpha}{2})}{\Gamma(\nu+\frac{1}{2})\Gamma(\nu+\frac{3}{2})} = 1+ \frac{\alpha(\alpha-2)}{4\nu} + o(1/\nu),
\end{equation}
so that it only remains to obtain the asymptotics of $f_{k}(\nu)$ defined in (\ref{eq:2}). This is the most difficult issue, which is explicitly solved in Appendix \ref{asymptotics:app} where we have found the first two terms of the asymptotics:
\begin{equation}
\label{eq:9}
f_{k}(\nu) = \frac{k!}{(2\nu)^{k}}\left(1-\frac{k(k+3+2\alpha(2-\alpha))}{4\nu} +\mathcal  O\left(\frac{1}{\nu^{2}}\right)  \right).
\end{equation}
Then, inserting (\ref{eq:11}), (\ref{eq:10}) and (\ref{eq:9}) in (\ref{eq:1bis}) we get 
\begin{equation}
\label{eq:12}
\langle p^{\alpha}\rangle =\frac{Z^{\alpha}}{\eta^{\alpha}}\left(1+\frac{\alpha(\alpha-2)(2k+1)}{4\nu}+o(1/\nu) \right), \quad \nu\rightarrow +\infty.
\end{equation}
From this expression and taking into account that $\eta=n+\frac{D-3}{2}$ and $\nu = l+\frac{D-1}{2}$, one can obtain the asymptotics (\ref{eq:avepa}) at the limit $D\rightarrow \infty$ for the momentum expectation values of the $D$-dimensional hydrogenic system. It is worth mentioning that the method (see Appendix \ref{asymptotics:app}) admits further refinement to obtain next terms of the asymptotic expansion of $\langle p^{\alpha}\rangle$.

Now let us consider the momentum logarithmic expectation value $\langle \log p \rangle$. From the general expression (\ref{eq:logmomexpec}) one has this quantity in the large $D$ limit is given by
\begin{equation}
\label{eq:avelogp}
\langle \log p \rangle = -\frac{4 n-2l-4}{D} - \log D +\log (2Z) + \mathcal{O}\left(\frac{1}{D^{2}}\right).
\end{equation}

Finally, for the circular states $(l = n-1)$ the general expressions (\ref{eq:avepa}) and (\ref{eq:avelogp}) supply the following  momentum radial and logarithmic expectation values
\begin{equation}
\label{eq:avepacirc}
\langle p^{\alpha} \rangle_{cs} = \left(\frac{2Z}{D} \right)^{\alpha}\left(1+\frac{\alpha(\alpha-2)(2n-1)}{2D} + \mathcal{O}(D^{-2})\right),
\end{equation}
and
\begin{equation}
\label{eq:avelogpcirc}
\langle \log p \rangle_{cs} = -\frac{1}{D}-\log \left(\frac{D}{2}\right)+\log (Z)+ \mathcal{O}\left(\frac{1}{D^{2}}\right) ,
\end{equation}
respectively. Moreover, from these expressions we can obtain the position and logarithmic expectation values for the ground state ($n=1$) of the $D$-dimensional hydrogenic system at the $D\rightarrow \infty$ limit.

\section{Expectation values of large-$D$ for Rydberg hydrogenic states }

In this section we compute the radial expectation values in position and momentum spaces for D-dimensional Rydberg hydrogenic states $(n,l,\{\mu\})$ when $D\gg 1$ and $n\gg1$, being $(l,\{\mu\})$ uniformly bounded. The final expressions are Eqs. (\ref{eq:lag8}) and (\ref{eq:aveinfp}) in the two reciprocal spaces, respectively.
\subsection{Position space}
We begin with the expression (\ref{eq:radexpec1bis}) of the position expectation value of an arbitrary $D$-dimensional hydrogenic state characterized by the hyperquantum numbers $(n,l,\{\mu\})$,
\begin{equation}
\label{eq:r1}
2\eta\left(\frac{2Z}{\eta}\right)^{\alpha}\langle r^{\alpha} \rangle = \int_{0}^{\infty}w_{\nu}(t)[\mathcal{\widehat{L}}^{(\nu)}_{k}(t)]^{2}t^{\alpha+1}\, dt,
\end{equation}
with $k=n-l-1$ and $\nu = 2l+D-2$. This integral converges for all values of $\alpha > -2l-D$. For convenience we make the linear change $ t= kx$, so that we can rewrite the previous expression as
\begin{equation}
\label{eq:r2}
2\eta\left(\frac{2Z}{\eta}\right)^{\alpha}\langle r^{\alpha} \rangle = k^{\alpha+1} \int_{0}^{\infty}x^{\nu}e^{-kx}[\mathcal{\hat{L}}^{(\nu)}_{k}(x)]^{2}x^{\alpha+1}\, dx,
\end{equation}
where the polynomial
\begin{equation}
\label{eq:r3}
\mathcal{\hat{L}}^{(\nu)}_{k}(x) \equiv k^{\frac{\nu+1}{2}}\mathcal{\widehat{L}}^{(\nu)}_{k}(kx)
\end{equation}
is orthonormal on $[0,+\infty)$ with respect to the weight $x^{\nu}e^{-kx}$. We want to determine its asymptotics when $n$ and $D$ tend simultaneously to infinity and  $l$ is uniformly bounded; that is, when both $k$ and $\nu$ tend  to infinity simultaneously. So,  $\lim_{n\to+\infty} \frac{l}{n} =0$ and we assume that
\begin{equation}
\label{eq:condknu1}
\lim_{k\to+\infty} \frac{\nu}{k} =\lambda \in (0+\infty).
\end{equation} 
In this situation,  polynomials $\mathcal{\hat{L}}^{(\nu)}_{k}(x)$ given by (\ref{eq:r3}) are orthogonal with respect to a varying weight, i.e. a weight which depends on the degree $k$ in the form 
\begin{equation}
\label{eq:r4}
w_{k}(x) = x^{\alpha_{k}}e^{-\beta_{k}x} \quad \text{with} \quad \alpha_{k} =\nu \quad \text{and} \quad \beta_{k} =k.
\end{equation}
It is a known fact (see \cite[Chap. 7]{saff}, also \cite{aptekarev}) that  the  modified Laguerre function converges in the weak-* sense, as $k\to \infty$:
\begin{equation}
\label{eq:lag1}
[\hat{\mathcal{L}}_{k,k}^{(\nu)}(x)]^{2}w_{k}(x)dx\rightarrow d\mu_{1} (x) := \frac{1}{\pi}\frac{\sqrt{(x-a)(b-x)}}{x}\, dx , \quad a < x < b ,
\end{equation}
where $\mu_1$ is the equilibrium measure on $\mathbb R_+$ in the external field
\begin{equation}
\label{eq:exfi}
\phi(x) = -\frac{\lambda}{2}\log x + \frac{x}{2};
\end{equation}
it is supported on the interval $[a,b]$ given explicitly by
\begin{equation}
\label{eq:lag2}
a=a_{\lambda} = \lambda + 1 - \sqrt{1+2\lambda} , \quad b=b_{\lambda} = \lambda + 1 +\sqrt{1+2\lambda} .
\end{equation}
Thus,
\begin{equation}
\label{eq:lag3}
\lim_{k\to+\infty}\int_{0}^{\infty} x^{\nu}e^{-kx}[\hat{\mathcal{L}}_{k,k}^{(\nu)}(x)]^{2} x^{\alpha+1}\, dx = \frac{1}{\pi } \int_{a_{\lambda}}^{b_{\lambda}} x^{\alpha} \sqrt{(x-a_{\lambda})(b_{\lambda}-x)}\, dx.
\end{equation}
With the change of variable $z=\frac{x-a_{\lambda}}{b_{\lambda}-a_{\lambda}}$ we get
\begin{eqnarray}
\label{eq:lag4}
\lim_{k\to+\infty}\int_{0}^{\infty} x^{\nu}e^{-kx}[\hat{\mathcal{L}}_{k,k}^{(\nu)}(x)]^{2} x^{\alpha+1}\, dx &=& \frac{a_{\lambda}^{\alpha}(b_{\lambda}-a_{\lambda})^{2}}{\pi}\int_{0}^{1} \left(1+z\frac{b_{\lambda}-a_{\lambda}}{a_{\lambda}}  \right)^{\alpha} z^{\frac{1}{2}}(1-z)^{\frac{1}{2}}\, dz \nonumber\\
&=& \frac{a_{\lambda}^{\alpha}(b_{\lambda}-a_{\lambda})^{2}}{8}\, _2F_{1}\left(-\alpha,\frac{3}{2}; 3; \frac{a_{\lambda}-b_{\lambda}}{a_{\lambda}}  \right),
\end{eqnarray}
where we have used the integral representation of the $_2F_{1}$,
\begin{equation}
\label{eq:lag5}
_2F_{1}(a,b;c;z)  = \frac{\Gamma(c)}{\Gamma(c-b)\Gamma(b)}\int_{0}^{1} t^{b-1}(1-t)^{c-b-1}(1-zt)^{-a}\, dt.
\end{equation}
by equation (\ref{eq:r2}) we obtain 
\begin{equation}
\label{eq:lag6}
\lim_{k\to\infty} 2\eta\left(\frac{2Z}{\eta} \right)\langle r^{\alpha} \rangle = \frac{k^{\alpha+1}a_{\lambda}^{\alpha}(b_{\lambda}-a_{\lambda})^{2}}{8}\, _2F_{1}\left(-\alpha,\frac{3}{2}; 3; \frac{a_{\lambda}-b_{\lambda}}{a_{\lambda}}  \right).
\end{equation}
Then, for $k\rightarrow +\infty$ and $\nu \rightarrow +\infty$ satisfying (\ref{eq:condknu1}) we finally get
\begin{eqnarray}
\label{eq:lag7}
\langle r^{\alpha} \rangle  &=& \frac{1}{2\eta}\left(\frac{\eta}{2Z} \right)^{\alpha}\frac{k^{\alpha+1}a_{\lambda}^{\alpha}(b_{\lambda}-a_{\lambda})^{2}}{8}\,_2F_{1}\left(-\alpha,\frac{3}{2}; 3; \frac{a_{\lambda}-b_{\lambda}}{a_{\lambda}}  \right)(1+o(1))\\
\label{eq:lag8}
&= & \frac{(2n+D)^{\alpha-1}n^{\alpha+1}}{2^{2\alpha+3}Z^{\alpha}}a^{\alpha}(b_{\lambda}-a_{\lambda})^{2}\, _2F_{1}\left(-\alpha,\frac{3}{2}; 3; \frac{a_{\lambda}-b_{\lambda}}{a_{\lambda}}  \right) (1+o(1)),
\end{eqnarray}
where in the second equality we have considered the approximations $k =n-l-1 \simeq n$ and $\eta = n + \frac{D-3}{2} \simeq n+ \frac{D}{2}$.
 
\subsection{Momentum space}

We turn to the general expression (\ref{eq:momint}) of the momentum expectation values $\langle p^{\alpha}\rangle$ of a generic $D$-dimensional hydrogenic state. Our aim is to determine the asymptotics of $\langle p^{\alpha}\rangle$ when $n$ and $D$ tend simultaneously to infinity and for $l$ uniformly bounded; that is, when both $k$ and $\nu$ tend  to infinity simultaneously and satisfy the condition (\ref{eq:condknu1}).\\

For convenience we rewrite (\ref{eq:momint}) as
\begin{eqnarray}
\label{eq:avmomentum}
 \langle p^{\alpha}\rangle &=& Z^{\alpha}\left(\frac{2}{2n+D-3}\right)^{\alpha} \int_{-1}^{1}(1-t)^{\alpha/2}(1+t)^{1-\alpha/2}[G_{k}^{(\nu)}(t)]^{2}c_{\nu}w^{\nu}(t)\, dt,
\end{eqnarray}
where the factor $c_{\nu}$ is given by
\begin{equation}
\label{eq:cnu}
c_{\nu} = \frac{\Gamma(\nu+1)}{\sqrt{\pi}\,\Gamma(\nu+1/2)}
\end{equation}
so that  
\begin{equation}
\label{eq:cnu2}
\int_{-1}^{1} c_{\nu}w^{\nu}(t)\, dt =1.
\end{equation}
The appropriately normalized Gegenbauer polynomials, 
\begin{equation}
\label{eq:gegenG}
G_{k}^{(\nu)}(x) = \left(\frac{k!(k+\nu)\Gamma(2\nu)}{\nu\Gamma(k+2\nu)}\right)^{1/2}\mathcal{C}_{k}^{(\nu)}(x),
\end{equation} 
are orthonormal with respect to the unit weight $c_{\nu}w^{\nu}(x)$, and exhibit the following weak-* asymptotics:
\begin{equation}
\label{eq:gegenG2}
[G_{k}^{(\nu)}(x)]^{2}c_{\nu}w^{\nu}(x) \, dx \longrightarrow \frac{1}{\pi}\frac{dx}{\sqrt{1-x^{2}}},\quad k\rightarrow \infty.
\end{equation}
Since $\nu,k\rightarrow +\infty$ with the condition (\ref{eq:condknu1}) satisfied, and
\begin{equation}
\label{eq:limit}
\lim_{k\to+\infty} -\frac{\log(c_{\nu}w_{\nu}(x))}{2k} =\frac{\lambda}{2}\log\frac{1}{1-x^{2}}, \quad x\in(-1,1),
\end{equation}
one has \cite{aptekarev} that the weak-* asymptotics of the orthonormal Gegenbauer polynomials $G_{k}^{(\nu)}(x)$ is given by
\begin{equation}
\label{eq:cond1}
[G_{k}^{(\nu)}(x)]^{2}c_{\nu}w^{\nu}(x)\, dx \longrightarrow d\mu_2(x), \quad k\rightarrow \infty,
\end{equation}
 on $[-1,1]$, where $\mu_2$ is the probability equilibrium measure on $[-1,1]$ in the external field
\begin{equation}
\label{eq:expot}
\phi(x) = \frac{\lambda}{2}\log\frac{1}{1-x^{2}}, \quad x\in (-1,1),
\end{equation}
created by two charges of size $\lambda/2$ fixed at $\pm 1$. The expression of $\mu_2$ is well-known (cf. \cite{saff}, Examples IV.1.17 and IV.5.2). It is supported on $[-\xi_{\lambda}, \xi_{\lambda}]$, with 
\begin{equation}
\label{eq:xi}
\xi_{\lambda} =\frac{\sqrt{\lambda+1/4}}{\lambda+1/2}>0,
\end{equation}
and
 \begin{equation}
 \label{eq:mup}
\mu_2'(x) = \left\{\begin{array}{cc}
 \frac{1+2\lambda}{\pi}\frac{\sqrt{\xi_{\lambda}^{2}-x^{2}}}{1-x^{2}} & \text{if} \,\, |x|\leq \xi_{\lambda}, \\
						 0 &  \text{otherwise}. \\
					\end{array}\right.
 \end{equation}
 Hence, we have 
 \begin{eqnarray}
 \label{eq:limint}
 \lim_{k\to\infty} \int_{-1}^{1} (1-t)^{\alpha/2}(1+t)^{1-\alpha/2}[G_{k}^{(\nu)}(x)]^{2}c_{\nu}w^{\nu}(t)\, dt &=& \lim_{k\to\infty} \frac{\Gamma(\nu+1)}{\sqrt{\pi}\Gamma(\nu+1/2)}\frac{k!(k+\nu)\Gamma(2\nu)}{\nu\,\Gamma(k+2\nu)}\nonumber\\
 & & \times \int_{-1}^{1} (1-t)^{\alpha/2}(1+t)^{1-\alpha/2}[\mathcal{C}_{k}^{(\nu)}(x)]^{2}w^{\nu}(t)\, dt \nonumber\\
 &=& \frac{1+2\lambda}{\pi}\int_{-\xi_{\lambda}}^{\xi_{\lambda}} (1-t)^{-1+\alpha/2}(1+t)^{-\alpha/2}\sqrt{\xi_{\lambda}^{2}-t^{2}}\, dt.\nonumber \\
 \end{eqnarray}
 From \cite{nist} we know that for $\text{Re}(\lambda) >0,\text{Re}(\mu)>0$, one has
 \begin{equation}
 \label{eq:intf1}
 \int_{0}^{1} x^{\lambda-1}(1-x)^{\mu-1}(1-ux)^{-\rho}(1-vx)^{-\sigma}\, dx = B(\mu,\lambda)\,F_{1}(\lambda,\rho,\sigma,\lambda+\mu;u,v),
 \end{equation}
 where  $B(x,y)=\Gamma(x) \Gamma(y) / \Gamma(x+y)$ is the Euler beta function, while the Appel's hypergeometric function $F_{1}(x,y)$ is defined for $|x|<1$, $|y|<1$ as
 \begin{equation}
 \label{eq:F1def}
 F_{1}(\alpha,\beta,\beta',\gamma;x,y) = \sum_{m=0}^{\infty}\sum_{n=0}^{\infty} \frac{(\alpha)_{m+n}(\beta)_{m}(\beta')_{n}}{(\gamma)_{m+n}m!n!}\,\,x^{m}\,y^{n},
 \end{equation}
 and extended analytically elsewhere. Since since $B(3/2,3/2) = \pi/8$,
 \begin{eqnarray}
 \label{eq:limint2}
 \int_{-\xi_{\lambda}}^{\xi_{\lambda}} (1-t)^{-1+\alpha/2}(1+t)^{-\alpha/2}\sqrt{\xi_{\lambda}^{2}-t^{2}}\, dt &=& 4\xi_{\lambda}^{2}(1-\xi_{\lambda})^{-1}\nonumber \\
 & & \times \int_{0}^{1} \left(1-\frac{2\xi_{\lambda}}{1+\xi_{\lambda}}x  \right)^{-1+\alpha/2}\left(1-\frac{-2\xi_{\lambda}}{1+\xi_{\lambda}}x\right)^{-\alpha/2}  x^{1/2}\sqrt{1-x}\,dx\nonumber\\
 &=& \frac{\pi}{8}\,F_{1}\left(\frac{3}{2},1-\frac{\alpha}{2},\frac{\alpha}{2},3;\frac{2\xi_{\lambda}}{1+\xi_{\lambda}},\frac{-2\xi_{\lambda}}{1+\xi_{\lambda}}\right).
 \end{eqnarray}
Thus, substituting in (\ref{eq:momint}), we get that for $k\rightarrow +\infty$ and $\nu\rightarrow +\infty$ (i.e. when $n$ and $D$ tend to infinity) satisfying $\lim_{k\to+\infty} \frac{\nu}{k} =\lambda \in (0+\infty)$, 
 \begin{eqnarray}
 \label{eq:aveinfp}
 \langle p^{\alpha}\rangle &=& Z^{\alpha}\left(\frac{2}{2n+D-3}\right)^{\alpha} \frac{1+2\lambda}{8} \,F_{1}\left(\frac{3}{2},1-\frac{\alpha}{2},\frac{\alpha}{2},3;\frac{2\xi_{\lambda}}{1+\xi_{\lambda}},\frac{-2\xi_{\lambda}}{1+\xi_{\lambda}} \right)\left(\frac{\Gamma(1+(D-1)/2)}{\Gamma(\nu)}\right)^{2}(1+o(1)),\nonumber  
 \end{eqnarray}
 where $\nu = 2l+D-2$.

\section{Uncertainty relations at the pseudoclassical limit}

In this section we study the uncertainty relations of Heisenberg and logarithmic types for the stationary states of a $D$-dimensional hydrogenic system at the pseudoclassical large-$D$ limit, and we illustrate that they fulfill the inequality-type uncertainty relations of both a general quantum system and  a system with a central potential. Let us advance that the Heisenberg-like uncertainty relation \cite{kennard}
\begin{equation}
\label{eq:UR1}
\langle r^{2}\rangle \langle p^{2}\rangle \geq \frac{D^{2}}{4},
\end{equation}
and the logarithmic-type uncertainty relation \cite{beckner}
\begin{equation}
\langle \log r \rangle + \langle \log p \rangle \ge \psi\left(\frac{D}{4}\right) +\log 2;
\quad l=0,1,2,\ldots
\label{eq:logrel1}
\end{equation}
are fulfilled for all stationary states of general $D$-dimensional quantum systems. Moreover, when the quantum-mechanical potential of the system is spherically symmetric, these uncertainty relations can be refined \cite{sanchez} as 
\begin{equation}
\label{eq:UR2}
\langle r^{2}\rangle \langle p^{2}\rangle \geq  \left(L+\frac{3}{2}\right)^{2} = \left(\frac{D}{2}+l\right)^{2}
\end{equation}
and \cite{beckner}
\begin{equation}
\langle \log r \rangle + \langle \log p \rangle \ge \psi\left(\frac{D+2l}{4}\right) +\log 2;
\quad l=0,1,2,\dots
\label{eq:logrel2}
\end{equation}
respectively.

\subsection{Heisenberg-like relations}

Taking into account the results of Section I, we have that the generalized Heisenberg-like uncertainty product $\langle r^{\alpha}\rangle \langle p^{\beta}\rangle$ of the $D$-dimensional hydrogenic system is given by
\begin{eqnarray}
\label{eq:radrel}
\langle r^{\alpha}\rangle \langle p^{\beta}\rangle &=& \frac{2 Z^k \eta ^{\alpha -k-1} \Gamma \left(\frac{1}{2} (D-k+2)+l\right) \Gamma \left(\frac{D+k}{2}+l\right) \Gamma (D+2 l+\alpha ) \Gamma (D+l+n-2)}{\left[\Gamma \left(\frac{D}{2}+l\right)\right]^2 \Gamma (D+2 l-1) \Gamma (D+2 l+1) \Gamma (n-l)}\nonumber\\
& & \times \, _3F_2(l-n+1,-\alpha -1,\alpha +2;1,D+2 l-1;1)\nonumber \\
& & \times \,  \, _5F_4\Bigg(\frac{D-1}{2}+l,\frac{1}{2} (D-k+2)+l,\frac{D+k}{2}+l,l-n+1,D+l+n-2 \nonumber \\
& & \quad\quad \quad;\frac{D}{2}+l,\frac{D+1}{2}+l,\frac{D}{2}+l+1,D+2 l-1;1 \Bigg),
\end{eqnarray}
\doubt which holds for $\alpha > -D-2l$ and $\beta \in (-D-2l, D+2l+2)$. Here again the notations $k = \eta + L +1 = n-l-1$ and $\nu = L+1 = l + (D-1)/2$ have been used.  Note that for for $\alpha = \beta$, the corresponding generalized uncertainty product $\langle r^{\alpha}\rangle \langle p^{\alpha}\rangle$ can be simplified further and, moreover, it does not depend on the nuclear charge $Z$ as one would expect. For the particular case $\alpha = \beta =2 $ this expression provides the following Heisenberg uncertainty product 
\begin{equation}
\label{eq:2alphabeta}
\langle r^{2}\rangle \langle p^{2}\rangle = \frac{D^{2}}{4}\left\{1+\frac{1}{D}(10n-6l-9)+\frac{1}{D^{2}}[10n(n-3)-6l(l-2)+20]  \right\},
\end{equation}
which fulfills not only the general uncertainty relation (\ref{eq:UR1}) but also the refined uncertainty relation (\ref{eq:UR2}), as it should. Note that for the ground state we have the exact uncertainty relationship
$$\langle r^{2}\rangle_{gs} \langle p^{2}\rangle_{gs} =\frac{D^{2}}{4}\left(1 + \frac{1}{D}\right)$$
 
What happens at the pseudoclassical large-$D$ limit? Taking into account (\ref{eq:averk1}) and (\ref{eq:avepa}), one has the following expression for the generalized Heisenberg-like uncertainty product of a general hydrogenic state $(n,l,\{\mu  \})$ at the large-$D$ limit:
 \begin{eqnarray}
 \label{eq:rpprod}
 \langle r^{\alpha}\rangle \langle p^{\beta}\rangle &=& \left(\frac{D^{2}}{4Z}\right)^{\alpha}\left(\frac{D}{2Z}\right)^{-\beta}\left(1+\frac{(\alpha+1)(\alpha+4l-2)}{2D} \right) \left(1+\frac{(\alpha +1) (\alpha +2) (n-l-1)}{D}\right)\nonumber\\
 & & \times \left(1+\frac{(\beta-2) \beta (2n-2l-1)}{2 D}\right)\left(1+o(1)\right),
 \end{eqnarray}
which holds for $\alpha > -D-2l$ and $\beta \in (-D-2l, D+2l+2)$.\\
 Particular case: for circular hydrogenic states $(n,n-1,\{ n-1 \})$ one obtains
 \begin{eqnarray}
 \label{eq:circrpprod}
 \langle r^{\alpha}\rangle_{cs} \langle p^{\beta}\rangle_{cs} &= &  \left(\frac{D^{2}}{4Z}\right)^{\alpha}\left(\frac{D}{2Z}\right)^{-\beta}\left(1+\frac{(\alpha+1)(4n-6)}{2D} \right) \left(1+\frac{\beta(\beta-2)}{2 D}\right)\left(1+o(1)\right),
 \end{eqnarray}
which for $\alpha = \beta =2 $ gives
 \begin{equation}
 \label{eq:alphabeta2}
 \langle r^{2}\rangle_{cs} \langle p^{2}\rangle_{cs} = \frac{D^{2}}{4} \left[1+\frac{6 (n-1)}{D}\right]\left(1+o(1)\right).
 \end{equation}
 Then, for the ground state ($n=1$) we have that $\langle r^{2}\rangle_{gs} \langle p^{2}\rangle_{gs} =\frac{D^{2}}{4}$, so obtaining the equality in the general Heisenberg lower bounds given by (\ref{eq:UR1}) and (\ref{eq:UR2}).
 
 \subsection{Logarithmic relations}
 
The logarithmic uncertainty relation of a $D$-dimensional hydrogenic system has the form
\begin{eqnarray}
\label{eq:logrel}
\langle \log r\rangle+ \langle \log p\rangle &=&\frac{2n-2l-1}{2n+D-3}+\frac{(2n+D-3)(2l+D-2)}{(2n+D-3)^{2}-1}-\log 2 -1 +\psi(n+1+D-2), \nonumber 
\end{eqnarray}
where $k=n-l-1$ and $\nu=L+1=l +\frac{D-1}{2}$. Note again that this uncertainty relation does not depend on the nuclear charge, as one could expect. Taking into account again that \cite{nist} $\psi(z) = \log z -\frac{1}{2z} + o(1/z)$ for $z \rightarrow  \infty$, one finds that at the large-$D$ limit, this relation gets refined as
 \begin{equation}
 \label{eq:sumrp}
 \langle \log r \rangle +\langle \log p \rangle = \log \frac{D}{2} + \frac{n+l-\frac{5}{2}}{D} + \mathcal{O}\left(\frac{1}{D^{2}}\right).
 \end{equation}
 Particular case: for the circular hydrogenic states which have $l=n-1$, one has that
 \begin{equation}
 \label{eq:circsumrp}
 \langle \log r \rangle_{cs} +\langle \log p \rangle_{cs} = \log \frac{D}{2} + \frac{2n-\frac{7}{2}}{D} + \mathcal{O}\left(\frac{1}{D^{2}}\right),
 \end{equation}
 so that for the ground state ($n=1$) one obtains
 $$ \langle \log r \rangle_{gs} +\langle \log p \rangle_{gs} = \log \frac{D}{2} -\frac{3}{2D} + \mathcal{O}\left(\frac{1}{D^{2}}\right),$$ 
 which saturates the general uncertainty inequalities (\ref{eq:logrel1}) and (\ref{eq:logrel2}).\\
 Let us finally highlight that the general uncertainty inequalities of the $D$-dimensional quantum systems (\ref{eq:UR1})-(\ref{eq:logrel2}) saturate, i.e. become uncertainty equalities, for the $D$-dimensional hydrogenic atom.
  
\section{Bounds on entropic uncertainty measures at large $D$}

The Shannon and Rényi entropies of general $D$-dimensional quantum systems not only describe numerous fundamental quantities of these systems but also characterize most appropriately uncertainty measures, both in position and momentum spaces. However, they cannot be computed in a closed form for the stationary states of the system, except for those lying at the two extremes of the associated energetic spectrum (particularly the ground state and the Rydberg states) of the hydrogenic \cite{yanez,dehesa10,lopez2013,toranzo16} and oscillator-like \cite{,aptekarev12,aptekarev15} systems.

  In this section we obtain upper bounds on the Shannon \cite{shannon} and Rényi \cite{renyi70} entropies and lower bounds on the Tsallis entropy \cite{tsallis88} of arbitrary stationary states of the $D$-dimensional hydrogenic states at the large $D$ limit in terms of $D$ and the states' hyperquantum numbers. The hydrogenic Shannon, Rényi and Tsallis entropies are defined by the following logarithmic and power functionals of the electron probability density $\rho(\vec{r}) \equiv \rho_{n,l,\{\mu\}}(\vec{r})$ : 
\begin{equation}\label{eq:shannon}
S[\rho]:= -\int_{R_D} \rho(\vec{r}) \log  \rho(\vec{r})  d \vec{r},
\end{equation} 
\begin{equation}\label{eq:renyi}
R_q[\rho]:=\frac{1}{1-q} \log W_q[\rho]=\frac{1}{1-q} \log \int_{R_D} \left[\rho(\vec{r})\right]^qd \vec{r}, 
\end{equation}
and
\begin{equation}\label{eq:tsallis}
T_q[\rho]:= \frac{1}{q-1} \left[ 1-W_q[\rho] \right]=\frac{1}{q-1} \left\{ 
1-\int_{R_D} \left[\rho(\vec{r})\right]^q d \vec{r}
 \right\}
\end{equation} 
respectively, where $\rho(\vec{r})$ is given by (\ref{eq:denspos}) and $q >0,\, q \ne 1$. Notice that when $q \rightarrow 1$, both Rényi and Tsallis entropies reduce to Shannon entropy.

\subsection{Upper bounds}
It is known \cite{lopez08} that the the Shannon entropy of general quantum systems have the following upper bounds, valid for all $\alpha>0$,
\begin{equation}
\label{eq:boundS2}
 S[\rho]\le A_0(\alpha,D)+\frac{D}{\alpha}\log{\left< r^\alpha \right>},
\end{equation}
with
\begin{equation}
	A_0(\alpha,D) = \frac{D}{\alpha}+\log \left[\frac{2\pi^{\frac{D}{2}}}{\alpha}\left(\frac{\alpha}{D}\right)^{\frac{D}{\alpha}}\frac{\Gamma(\frac{D}{\alpha})}{\Gamma(\frac{D}{2})} \right].
\end{equation}
 The Rényi entropy $R_q[\rho]$ can be bounded from above in terms of $\langle r^\alpha \rangle$, with $\alpha\in \mathbb N$, by
\begin{equation}
\label{eq:boundRk}
R_q[\rho]\le \frac{1}{1-q}\log \left\{ L_1(q,\alpha,D) \left\langle r^\alpha\right\rangle^{-\frac{D}{\alpha}(q-1)}\right\}
\end{equation}
and in terms of $\left\langle r^{-\alpha} \right\rangle$, with $\alpha\in \mathbb N$, but subject to the 
condition $\alpha < \frac{D}{q} (q-1)$, by
\begin{equation}\label{eq:boundRmk}
R_q[\rho]\le \frac{1}{1-q}\log \left\{L_2(q,\alpha,D) \left\langle r^{-\alpha}\right\rangle^{-\frac{D}{\alpha}(q-1)}\right\}.
\end{equation}
Functions $L_i \left(q,\alpha,D \right)$, $i=1,2$, have an explicit expression,
\begin{equation}\label{eq:funcL1}
L_1 \left( q,\alpha,D \right)=\frac{q\alpha}{D (q-1)+\alpha q} \left\lbrace \frac{\alpha \Gamma \left(D/2 \right) \left[ 
\frac{D(q-1)}{D(q-1)+\alpha q} \right]^{\frac{D}{\alpha}}}{2 \pi^{\frac{D}{2}} B \left(\frac{q}{q-1},\frac{D}{\alpha} 
\right)} \right\rbrace^{q-1} 
\end{equation}
and
\begin{equation}\label{eq:funcL2}
L_2 \left( q,\alpha,D \right)=\frac{q\alpha}{D (q-1)-\alpha q} \left\lbrace \frac{\alpha \Gamma \left(D/2 \right) \left[
 \frac{D(q-1)-\alpha q}{D(q-1)} \right]^{\frac{D}{\alpha}}}{2 \pi^{\frac{D}{2}} B \left(\frac{D}{\alpha}-\frac{1}{q-1},
\frac{q}{q-1} \right)} \right\rbrace^{q-1} .
\end{equation}
To derive these upper bounds we have used the variational bounds \cite{dehesa88} on the entropic moments $W_\alpha[\rho]$ with a single expectation value $\langle r^\alpha \rangle$ as constraint. \doubt  At the pseudoclassical limit we have that
\begin{equation}
\label{eq:lS1}
A_{0}(\alpha,D) = - \frac{D-1}{2}\log \frac{D}{2} +\log (\pi e)\frac{D}{2} -\frac{1}{2} \log \frac{D}{\alpha} +\log \frac{2}{\alpha} +o(1),
\end{equation}
where we have used the asymptotic expansion \cite{nist} $\log \Gamma(z) = \left(z-\frac{1}{2}\right)\log z -z +\frac{1}{2}\log 2\pi+o(1)$ for  $z\rightarrow \infty$, and \doubt 
\begin{eqnarray}
\label{eq:lS2}
\frac{D}{\alpha}\log \langle r^{\alpha}\rangle 
& \sim & 2D\log\frac{D}{2}-D\log Z + \frac{D}{\alpha} A_{1}(\alpha,D) ,
\end{eqnarray}
where the term $A_{1}(\alpha,D)$ is given by 
\begin{equation}
\label{eq:lS22}
A_{1}(\alpha,D)=\log\left[\left(1+\frac{(\alpha+1)(\alpha+4l-2)}{2D} \right)\left(1+\frac{(\alpha +1) (\alpha +2) (n-l-1)}{D}\right)  \right]
\end{equation}
and tends to $0$ as $D\rightarrow \infty$. These asymptotic approximations allow us to write the following inequalities  for the radial Shannon entropy \doubt 
\begin{equation}
\label{eq:lS3}
S[\rho] \lesssim A_{2}(\alpha,D) + \frac{D}{\alpha}A_{1}(\alpha,D) -D\log Z +\log \frac{2}{\alpha} ,
\end{equation}
where
\begin{eqnarray}
\label{eq:lS4}
A_{2}(\alpha,D) &=& 3D\log D + \log \left(\frac{\pi e}{8}\right)\frac{D}{2}  + \frac{1}{2} \log \frac{\alpha}{2}.
\end{eqnarray}
Rearranging all terms in (\ref{eq:lS3}) and (\ref{eq:lS4}) we can write the asymptotics of the upper bound for the Shannon entropy as
\begin{equation}
\label{eq:lS5}
S[\rho] \lesssim  3D\log D +\left[\log\left(\frac{\pi e}{8}\right)^{\frac{1}{2}}-\log Z + \frac{1}{\alpha}A_{1}(\alpha,D)\right]D -\frac{1}{2}\log\frac{\alpha}{2},
\end{equation}
at the large $D$ limit. Operating in a similar way for the radial Rényi entropy at the large-$D$ limit we find from (\ref{eq:boundRk})    that
\begin{eqnarray}
\label{eq:lR1}
R_{q}[\rho]  &\lesssim & \frac{1}{1-q}\log L_{1}(q,\alpha,D) +2D\log\frac{D}{2} - D\log Z + \frac{D}{\alpha}A_{1}(\alpha,D)\nonumber \\
&\lesssim & \frac{3D-1}{2}\log D +\left[ \log \left(\frac{\pi e}{8}\right)^{\frac{1}{2}} - \log Z+ \frac{1}{\alpha}A_{1}(\alpha,D)\right]D+\frac{1}{1-q}A_{3}(q),
\end{eqnarray}
where
\[
A_{3}(q) = \log \frac{q}{q-1} + (1-q)\log \Gamma\left(\frac{q}{q-1}\right) + \frac{1-q}{2}\log \frac{2}{\pi}.
\]
Since $\log L_{1} = \log L_{2}$ as $D\rightarrow \infty$, we obtain a similar expression for the lower bound in (\ref{eq:boundRmk}) by just changing $A_{1}(\alpha,D)$ to $A_{1}(-\alpha,D)$. 

\subsection{Lower bounds}
    
We know from \cite[Eqs. (1.49) and (1.50)]{dehesa2012} that the following inequalities for the Tsallis entropy hold: in terms of $\left\langle r^\alpha \right\rangle$ with $\alpha\in \mathbb N$,
\begin{equation}\label{eq:boundTqk}
1+(1-q)T_q\left[\rho \right] \geq L_1\left(q,\alpha,D \right)\left\langle r^\alpha \right\rangle^{-\frac{D}{\alpha}(q-1)}
\end{equation}
 and in terms of $\left\langle r^{-\alpha} \right\rangle$ with $\alpha\in \mathbb N$ and for $\alpha<\frac{-D(q-1)}{q}$, 
\begin{equation}\label{eq:boundTqmk}
1+(1-q)T_q\left[\rho \right] \geq L_2\left(q,\alpha,D \right)\left\langle r^{-\alpha} \right\rangle^{-\frac{D}{\alpha}(q-1)}.
\end{equation}
The functions $L_i \left(q,\alpha,D \right)$ are defined in Eqs. (\ref{eq:funcL1}) and (\ref{eq:funcL2}), respectively. Operating as in the previous subsection we find in the limit $D\rightarrow\infty$ the following asymptotic  lower bounds:
\begin{equation}
\label{eq:lT1}
1+(1-q)T_{p}[\rho] \gtrsim A_{5}(\alpha,D)A_{1}(\alpha,D)^{-\frac{D}{\alpha}(q-1)}
\end{equation}
and 
\begin{equation}
\label{eq:lT2}
1+(1-q)T_{p}[\rho] \gtrsim A_{5}(\alpha,D)A_{1}(-\alpha,D)^{-\frac{D}{\alpha}(q-1)},
\end{equation}
where 
\begin{equation}
\label{eq:lT3}
A_{5}(\alpha,D) = \frac{q}{q-1}\left[\frac{D^{3 D-1}}{2^{3 D+2}\pi ^{D+1}Z^{2D}}\right]^{\frac{1-q}{2}}\frac{e^{(1-q)\frac{D}{2} }}{\alpha^{q}}\Gamma \left(\frac{q}{q-1}\right)^{1-q}.
\end{equation}
Finally, let us comment that expressions similar to the inequalities (\ref{eq:boundS2}), (\ref{eq:boundRk}), (\ref{eq:boundRmk}), (\ref{eq:lT1}) and (\ref{eq:lT2}) for the position Shannon, Rényi and Tsallis entropies given by (\ref{eq:shannon})-(\ref{eq:boundS2}) are also valid for the corresponding quantities in the momentum space.

\section{Conclusions}

The main prototype of the $D$-dimensional Coulomb many-body systems, the $D$-dimensional hydrogenic system, is investigated by means of the radial expectation values in both position and momentum spaces. These expectation values, which characterize numerous fundamental and/or experimentally accesibles quantities of the system (e.g., kinetic and repulsion energies, diamagnetic susceptibility,\dots) and describe generalized Heisenberg-like uncertainty measures, are calculated for all quantum states of the system at the (pseudoclassical) large $D$ limit. Then, the uncertainty equality-type relations associated to them are determined, and show that they fulfill and saturate the known uncertainty inequality-type relations for both general quantum systems and for those systems with a quantum-mechanical spherically symmetric potential. Moreover, the position and momentum expectation values are used to bound the entropic uncertainty measures of the Shannon, Rényi and Tsallis types at large $D$. Finally, let us point out an open problem which is important \textit{per se}: the determination of these three entropies at this pseudoclassical limit for all quantum $D$-dimensional hydrogenic states, which is left for future work.

\section*{Acknowledgments}
The first and third authors were partially supported by Projects FQM-7276 and FQM-207 from Junta de Andaluc\'ia and by the Spanish Government together with the European Regional Development Fund (ERDF) under grants FIS2011-24540, FIS2014-54497 and FIS2014-59311-P. The first author additionally acknowledges the support of the Ministry of Education of Spain under the program FPU.

The second author was partially supported by the Spanish Government together with the European Regional Development Fund (ERDF) under grants MTM2011-28952-C02-01 (from MICINN) and MTM2014-53963-P (from MINECO), by Junta de Andaluc\'{\i}a (the Excellence Grant P11-FQM-7276 and the research group FQM-229), and by Campus 
de Excelencia Internacional del Mar (CEIMAR) of the University of Almer\'{\i}a. 

\appendix

\section{Asymptotics of $f_{k}(\nu)$}
\label{asymptotics:app}

Here obtain in full detail the asymptotics of the quantities $f_{k}(\nu)$ defined in Eq. (\ref{eq:2}); that is,
\begin{equation}
f_{k}(\nu) = \frac{1}{(2\nu)_{k}}\sum_{j=0}^{k}(-1)^{j}\binom{k}{j}(2\nu+j)_{k}\,d_{j}
\end{equation}
where
\begin{eqnarray}
\label{eq:3}
d_{j}\equiv d_{j}(\nu) &=& \frac{\nu}{\nu+j}\frac{(\nu+\frac{\alpha+1}{2})_{j}(\nu+\frac{3-\alpha}{2})_{j}}{(\nu+\frac{1}{2})_{j}(\nu+\frac{3}{2})_{j}}\nonumber\\
&=& \frac{\nu}{\nu+j}\prod_{i=1}^{j}\left(1-\frac{p}{(\nu+i+\frac{1}{2})(\nu+i-\frac{1}{2})}\right),
\end{eqnarray}
and $p \equiv p(\alpha ) = \frac{1}{4}\alpha(\alpha-2)$.
We first establish two technical results (Lemma 1 and Proposition 1), which allow us to express  the quantities $f_{k}(\nu)$ in terms of the backward-difference operator $\nabla d_{k} = d_{k} - d_{k-1}$. Then, we derive the asymptotic expansions of $d_{k}$ and $\nabla^{i}d_{k}$ by means of Lemma 2 and Corollary 1, respectively. Finally, the Corollary 2 yields the wanted asymptotics of $f_{k}(\nu)$ at large $D$.\\

\textbf{Lemma 1}. For $0\leq j \leq k$,
\[
\frac{(a+j)_{k}}{(a)_{k}} = k! \sum_{i=0}^{j}\binom{j}{i}\frac{1}{(k-i)!(a_{i})}.
\]
\textbf{Proof}. Since
\begin{equation}
\label{eq:1a}
\frac{(a+j)_{k}}{(a)_{k}} = \frac{(a+k)_{j}}{(a)_{j}},
\end{equation}
the identity 
\[
(a+b)_{k} = \sum_{i=0}^{k}(-1)^{i}\binom{k}{i}(a+i)_{k-i}(-b)_{i}
\]
yields
\[
\frac{(a+k)_{j}}{(a)_{j}} = \sum_{i=0}^{j}\binom{j}{i}\frac{k!}{(k-i)!}\frac{(a+i)_{j-i}}{(a)_{j}}=k!\sum_{i=0}^{j}\frac{1}{(k-i)!(a)_{i}},
\]
where we have used that $(a+i)_{j-i} = \frac{(a)_{j}}{(a)_{i}}. \hspace{7cm} \square$\\

\textbf{Proposition 1}. For $f_{k}(\nu)$ given in (\ref{eq:2}) we have that it can be rewritten in the form
\begin{equation}
\label{eq:5}
f_{k}(\nu) = (-1)^{k}k!\sum_{i=0}^{k}\binom{k}{i}\frac{\nabla^{i}d_{k}}{i!(2\nu)_{k-i}},
\end{equation}
where $\nabla$ denotes the operator of backward difference, i.e.,
\[
\nabla d_{k} = d_{k} - d_{k-1}, \quad \nabla^{n+1}d_{k} = \nabla(\nabla^{n}d_{k}).
\]
\textbf{Proof}. By Lemma 1 we have
\begin{eqnarray*}
\label{eq:6a}
f_{k}(\nu) &=& \sum_{j=0}^{k}(-1)^{j}\binom{k}{j}d_{j}\frac{(2\nu+j)_{k}}{(2\nu)_{k}} \\
&=& \sum_{j=0}^{k} (-1)^{j}\binom{k}{j}d_{j}\left(k!\sum_{i=0}^{j}\binom{j}{i}\frac{1}{(k-i)!(2\nu)_{i}}\right)\\
&=& k!\sum_{j=0}^{k}\sum_{i=0}^{j}(-1)^{j}\binom{k}{j}\binom{j}{i}d_{j}\frac{1}{(k-i)!(2\nu)_{i}}\\
&=&  k!\sum_{i=0}^{k}\sum_{j=i}^{k}(-1)^{j}\binom{k}{j}\binom{j}{i}d_{j}\frac{1}{(k-i)!(2\nu)_{i}}\\
&=& k!\sum_{i=0}^{k}\sum_{j=0}^{k-i}(-1)^{j+i}\binom{k}{j+i}\binom{j+i}{i}d_{j+i}\frac{1}{(k-i)!(2\nu)_{i}}.
\end{eqnarray*}
Since
\[
\binom{k}{j+i}\binom{j+i}{i} = \binom{k-i}{j}\binom{k}{i},
\]
we obtain
\[
f_{k}(\nu) = k!\sum_{i=0}^{k}\binom{k}{i}\frac{1}{(k-i)!(2\nu)_{i}}\left(\sum_{j=0}^{k-i}(-1)^{j}\binom{k-i}{j}d_{j+i}\right).
\]
It remains to observe that
\[
\sum_{j=0}^{n}(-1)^{j}\binom{n}{j}d_{k-j} = \nabla^{n}d_{k},\quad n=0,\ldots,k,
\]
so that
\[
\sum_{j=0}^{k-i}(-1)^{j}\binom{k-i}{j}d_{j+i} =(-1)^{k-i} \nabla^{k-i}d_{k}. \qquad \qquad \square
\]

\textbf{Lemma 2}. The asymptotic expansion of $d_{k} = d_{k}(\nu)$ is given by
\begin{equation}
\label{eq:6}
d_{k} = d_{k}(\nu) = \sum_{n=0}^{\infty} (-1)^{n}\frac{\beta_{n}(k)}{\nu^{n}},\quad \nu\rightarrow +\infty,
\end{equation}
where $\beta_{n}(k)$ are monic polynomials in $k$. Furthermore, $\beta_{0}(k) = 1$ and for $n\geq 1$,
\begin{equation}
\label{eq:7}
\beta_{n}(k) = k^{n}-(n-1)p(\alpha)k^{n-1} + \text{lower degree terms}.
\end{equation}
\textbf{Proof}. We prove the result by induction in $k$. Observe first that $d_{0} = 1$ and for $k\geq 1$,
\begin{eqnarray*}
\frac{d_{k}}{d_{k-1}} &=& 1- \frac{4p+1}{k+\nu} + \frac{p}{\nu+k-1/2} + \frac{3p}{\nu+k+1/2}\\
&=& 1+ \sum_{n=1}^{\infty} (-1)^{n}\frac{(4p+1)k^{n-1}-p(k-1/2)^{n-1}-3p(k+1/2)^{n-1}}{\nu^{n}}\\
&=& \sum_{n=0}^{\infty}(-1)^{n}\frac{\gamma_{n}(k)}{\nu^{n}},
\end{eqnarray*}
where
\[
\gamma_{0}(k) = 1 \quad \text{and} \quad  \gamma_{n}(k) = k^{n-1}-p(n-1)k^{n-2}+\ldots \quad \text{for} \quad n\geq 1.
\]
Therefore, by assumptions,
\[
d_{k} =\left(\sum_{m=0}^{\infty}(-1)^{m}\frac{\beta_{m}(k-1)}{\nu^{m}}\right)\left(\sum_{n=0}^{\infty}(-1)^{n}\frac{\gamma_{n}(k)}{\nu^{n}}\right) = \sum_{r=0}^{\infty}(-1)^{r}\frac{\zeta_{r}(k)}{\nu^{r}},
\]
where
\[
\zeta_{r}(k) = \sum_{m+n=r}\beta_{m}(k-1)\gamma_{n}(k).
\]
For $r\geq 1$ we have
\begin{eqnarray*}
\zeta_{r}(k) &=& \beta_{r}(k-1)+\gamma_{r}(k) + \sum_{\substack{m+n=r\\ 0<m,n<r}} \beta_{m}(k-1)\gamma_{n}(k)\\ &=& k^{r}-(p+1)(r-1)k^{r-1} + \ldots\\
& & + \sum_{\substack{m+n=r\\ 0<m,n<r}} \{(k-1)^{m}-p(m-1)(k-1)^{m-1}+\ldots\}\{k^{n-1}-p(n-1)k^{n-2}+\ldots \} \\
&=&  \sum_{m+n=r}  \{k^{m} - (pm -p+m)k^{m-1} + \ldots   \} \{ k^{n-1} - p(n-1)k^{n-2}+ \ldots   \}\\
&=& k^{r} - (p+1)(r-1)k^{r-1} + \left(\sum_{\substack{m+n=r\\ 0<m,n<r}} k^{r-1}  \right) + \text{lower degree terms} \\
&=& k^{r} -(p+1)(r-1)k^{r-1}+(r-1)k^{r-1} + \text{lower degree terms} \\
&=& k^{r} -p(r-1)k^{r-1} + \text{lower degree terms},
\end{eqnarray*}
and the assertion follows. \hspace{9cm } $\square$\\

\textbf{Corollary 1}: The asymptotic expansion of $\nabla^{n} d_{k}$ is given by
\begin{equation}
\label{eq:8}
\nabla^{n} d_{k} = \frac{(-1)^{n}n!}{\nu^{n}}\left(1+\frac{np-(n+1)(k-n/2)}{\nu} +\mathcal O(\nu^{-2})\right), \quad \nu\rightarrow +\infty,
\end{equation}
where $p \equiv p(\alpha ) = \frac{1}{4}\alpha(\alpha-2)$.\\
\textbf{Proof}. It is sufficient to observe that $\nabla^{n}k^{r} = 0$ for $r<n$,
\[
\nabla^{n}k^{n} = (-1)^{n}n! \quad \text{and} \quad \nabla^{n}k^{n+1} = (-1)^{n} n!(k-n/2). \qquad \qquad\square
\]
\textbf{Corollary 2}: The asymptotic expansion of $f_{k}(\nu)$ is given by
\begin{equation}
f_{k}(\nu) = \frac{k!}{(2\nu)^{k}}\left(1-\frac{k(k+3+2\alpha(2-\alpha))}{4\nu} + \mathcal O\left(\frac{1}{\nu^{2}}\right)  \right).
\end{equation}
\textbf{Proof}.  Just use Corollary 1 in (\ref{eq:5}).

\section{Table of convergence of the asymptotics of $\langle r^{\alpha}\rangle$ and $\langle p^{\alpha}\rangle$ }
 \label{table:app}
 
In this table it is shown the rate of convergence for the position and momentum expectation values of the $D$-dimensional hydrogen state ($n=2, l=0$) at large $D$ to the known exact values given in (\ref{eq:averk1}), (\ref{eq:avepa}), (\ref{eq:radexpec1bis}) and (\ref{eq:momint}), respectively.
\begin{table}[H]
\centering 
\setlength{\tabcolsep}{12pt}
\begin{tabular}{c c c c c c } 
\hline 
 $D$ & $\alpha$ & $\langle r^{\alpha}\rangle_{\text{asymp}}$ & $\langle p^{\alpha}\rangle_{\text{asymp}}$ & $\langle r^{\alpha}\rangle_{\text{exact}}$ & $\langle p^{\alpha}\rangle_{\text{exact}}$
\\ [0.5ex]
\hline 
50 & &1.00199  & 1. & & \\
250 & 0 & 1.00199  & 1. & 1 & 1 \\
500 &  & 1.00199 & 1.  & &  \\\hline
50 & &  686. &0.0388 & 612.5 &  0.0380789 \\
250 & 1 & 15936.. & 0.007952 & 15562.5 & 0.00792065   \\
500 &  & 63123.5. & 0.003988 & 62375. &  0.00398008 \\\hline
50 & & 484375. & 0.0016 & 365766. & 0.00153787 \\
250 & 2 &$2.55859\cdot 10^8$  & 0.000064 & $2.41176\cdot 10^8$  & 0.0000634911 \\
500 &  & $4\cdot 10^9$  & 0.000016 & $3.88267\cdot 10^9$  & 0.0000159362 \\\hline
50 & & 0.0016 & 27.25 & 0.00160064 & 27.7927 \\
250 & -1 & 0.000064 & 127.25 & 0.000064001 & 127.758 \\
500 &  & 0.000016 & 252.25 & 0.0000160001 & 252.754 \\\hline
\end{tabular}
\caption{Rate of convergence of the asymptotic expectation values in terms of $D$ for the hydrogen ($Z=1$) state with $n=2$ and $l=0$.} 
\label{tab:PPer}
\end{table}


\begin{thebibliography}{99}
 
\bibitem{witten} E. Witten, Phys. Today 33 (1980) 38–43
\bibitem{yaffe1} L.G. Yaffe, Large N limits as classical mechanics, Rev. Mod. Phys. 54 (1982) 407
\bibitem{herschbach} D. R. Herschbach, J. Avery and O. Goscinski (eds), Dimensional Scaling in Chemical Physics (Kluwer Acad. Publ., London, 1993)
\bibitem{tsipis} C. T. Tsipis, V. S. Popov, D. R. Herschbach, and J. S. Avery, New Methods in Quantum Theory (Kluwer Academic Publishers, Dordrecht, 1996)
\bibitem{svidzinsky} A. Svidzinsky, G. Chen, S. Chin, M. Kim, D. Ma, R. Murawski, A. Sergeev, M. Scully and D. Herschbach. Bohr model and dimensional scaling analysis of atoms and molecules, Int. Rev. Phys. Chem. 27 (2008) 665–723
\bibitem{chatterjee} A. Chatterjee, Phys. Rep. 186, 249 (1990).
\bibitem{avery} J. Avery, Hyperspherical Harmonics and Generalized Stur- mians; Kluwer: Dordrecht, 2000.
\bibitem{dong} S.H. Dong, Wave Equations in Higher Dimensions. Springer, New York, 2011
\bibitem{krenn} M. Krenn, M. Huber, R. Fickler, R. Lapkiewicz, S. Ramelowa and A. Zeilinger, Generation and confirmation of a (100 × 100)-dimensional entangled quantum system, Proceedings of the National Academy of Sciences 111 (2014) 6243-6247
\bibitem{bellomo} G. Bellomo, A.R. Plastino and A. Plastino, Quantum state space-dimension as a quantum resource, Preprint 2015.
\bibitem{crann} J. Crann, D. W. Kribs, R. H. Levene and I. G. Todorov, Private algebras in quantum information and infinite dimensional complementarity,  arXiv:1510.06672 [quant-ph], 22 October 2015.
\bibitem{bender1} C. M. Bender, S. Boettcher, and L. R. Mead, J. Math. Phys. 5, 368 (1994).
\bibitem{bender2} C. M. Bender, S. Boettcher, and M. Moshe, J. Math. Phys. 5, 4941 (1994)
\bibitem{beldjenna} A. Beldjenna, J. Rudnick, and G. Gaspari, J. Phys. A 24, 2131 (1991).
\bibitem{bender3} C.M. Bender and K.A. Milton, Scalar Casimir effect for a D-dimensional sphere, Phys. Rev. D 50 (1994) 6547.
\bibitem{yaffe2} L. G. Yaffe, Phys. Today 36 (8), 50 (1983).
\bibitem{herschbach_87} D. R. Herschbach, J. Chem. Soc. Faraday Disc. 84, 465 (1987).
\bibitem{herschbach96} D. R. Herschbach, Int. J. Quantum Chem. 57, 295 (1996)
\bibitem{herschbach_2000} D. R. Herschbach, Annu. Rev. Phys. Chem. (2000)1–39
 \bibitem{herschbach86} D. R. Herschbach, J. Chem. Phys. 84 (1986) 838.
\bibitem{pasternack} S. Pasternack, Proc. Natl Acad. Sci. USA 23 (1938) 91
 \bibitem{ray} A. Ray, K. Mahata and P.P. Ray, Moments of probability distribution, wavefunctions, and their derivatives at the origin of N-dimensional central potentials, Am. J. Phys. 56 (1988) 462
\bibitem{drake}  G.W.F. Drake and R.A. Swainson, Phys. Rev. A 42 (1990) 1123
 \bibitem{andrae} Andrae D 1997 J. Phys. B: At. Mol. Opt. Phys. 30 4435
 \bibitem{tarasov} Tarasov V F 2004 Int. J. Mod. Phys. B 18 3177–84
\bibitem{guerrero11} A. Guerrero, P. Sánchez-Moreno and J.S. Dehesa, Phys. Rev. A 84, 042105 (2011)
 \bibitem{hey2} J.D. Hey, Am. J. Phys. 61(1993) 28
 \bibitem{assche} W. van Assche, R.J. Yáñez, R. González-Férez and J.S. Dehesa, J.Math.Phys.41 (2000) 6600
 \bibitem{dehesa12} J.S. Dehesa, S.López-Rosa, P. Sánchez-Moreno and R. J. Yáñez, Complexity of multidimensional hydrogenic systems, Int. J. Appl. Math. Stat. 26 (2012) 973
 \bibitem{dehesa10} J. S. Dehesa, S. López-Rosa, A. Martínez-Finkelshtein, R. J. Yáñez, Information Theory of $D$-Dimensional Hydrogenic Systems: Application to Circular and Rydberg States, Int. J.  Quantum Chemistry 110, 1529–1548 (2010).
 \bibitem{zozor11} S. Zozor, M. Portesi, P. Sánchez-Moreno and J.S. Dehesa, Phys. Rev. A83 (2011) 052107
 \bibitem{toranzo16} I.V. Toranzo, S. López-Rosa, R.O. Esquivel and J.S. Dehesa, J. Phys. A: Math. Theor. 49 (2016) 025301
  \bibitem{aptekarev} A. I. Aptekarev, J. S. Dehesa, A. Martínez-Finkelshtein and R. J. Yáñez, Quantum expectation values of $D$-dimensional Rydberg hydrogenic states by use of Laguerre and Gegenbauer asymptotics, J. Phys. A: Math. Theor. 43 (2010) 145204 (10pp)
  \bibitem{knotterus} U. J. Knottnerus, Approximation Formulae for Generalized Hypergeometric Functions for Large Values of the Parameters. Groningen: J. B. Wolters, 1960.
  \bibitem{luke} Y. L. Luke (1969b). The Special Functions and their Approximations. Vol. 2. New York: Academic Press.
 \bibitem{nist} F. W.J. Olver, D.W. Lozier, R.F. Boisvert and C.W. Clark, NIST Handbook of Mathematical Functions (Cambridge University Press, New York, 2010).
\bibitem{nieto} M.M. Nieto, Am. J. Phys. 47 (1979) 1067
\bibitem{yanez} Yáñez, R. J., van Assche, W. and Dehesa, J. S., Position and momentum information entropies of the D-dimensional harmonic oscillator and hydrogen atom, Phys. Rev. A 50 (1994) 3065.
\bibitem{aquilanti} V. Aquilanti, S. Cavalli and C. Coletti, Chem. Phys. 214 (1997) 1–13
 \bibitem{szmytkowski} R. Szmytkowski, Solution of the momentum-space Schrödinger equation for bound states of the N-dimensional Coulomb problem (revisited), arXiv:1111.1661v1 [quant-ph] 7 Nov 2011
 \bibitem{delbourgo} R. Delbourgo and D. Elliott, Inverse momentum expectation values for hydrogenic systems, J. Math. Phys. 50 (2009) 062107 
 \bibitem{saff} E. B. Saff, V. Totik, \textit{Logarithmic Potentials with External Fields}, vol. 316 of \textit{Grundlehren der Mathematischen Wissenschaften}. Springer-Verlak, Berlin, 1997.
 \bibitem{kennard} Kennard E H 1927 Z. Phys. 44 326
 \bibitem{beckner} W. Beckner. Pitt’s inequality and the uncertainty principle. Proceed. Amer. Math. Soc., 123:1897–1905, 1995.
 \bibitem{rudnicki} Ł. Rudnicki, Pablo Sánchez-Moreno and J.S. Dehesa, J. Phys. A: Math. Theor. 45 (2012) 225303 (11pp)
 \bibitem{sanchez} Sánchez-Moreno, P.; González-Férez, R.; Dehesa, J. S., New J Phys 2006, 8, 330.
 \bibitem{lopez2013} S. López-Rosa, I. V. Toranzo, P. Sánchez-Moreno and J. S. Dehesa, J. Math. Phys. 54, 052109 (2013)
 \bibitem{toranzo16} I.V. Toranzo and J.S. Dehesa, EPL (2016). Submitted
 \bibitem{aptekarev12} A. I. Aptekarev, J. S. Dehesa, P. Sánchez-Moreno and D. N. Tulyakov, Contemp. Math. 578, 19-29 (2012)
 \bibitem{aptekarev15} A.I. Aptekarev, D. N. Tulyakov, I.V. Toranzo and J.S. Dehesa, Europ. Phys. J. B (2016). Accepted
 \bibitem{shannon} C. E. Shannon. A mathematical theory of communication. Bell Syst. Tech. J., 27:379, 1948.
 \bibitem{renyi70} A. Rényi. Probability Theory. Academy Kiado, Budapest, 1970.
 \bibitem{tsallis88} C. Tsallis. Possible generalization of Boltzmann-Gibbs statistics. J. Stat. Phys., 52:479, 1988.
 \bibitem{dehesa88} J. S. Dehesa and F. J. Galvez. Rigorous bounds to density-dependent quantities of D-dimensional many-fermion systems. Phys. Rev. A, 37:3634, 1988
 \bibitem{lopez08} S. López-Rosa, J.C.Angulo, J.S.Dehesa and R.J.Yáñez, Physica A, 387:2243–2255, 2008. Erratum, ibid 387, 4729-4730 (2008).
 \bibitem{dehesa2012} J.S.Dehesa, S.López-Rosa and D.Manzano, in K. D. Sen(ed.), Statistical Complexities: Application to Electronic Structure (Springer, Berlin, 2012). 
 \bibitem{loeser} J. G. Loeser, in chapter 9 of  D. R. Herschbach, J. Avery and O. Goscinski (eds), Dimensional Scaling in Chemical Physics (Kluwer Acad. Publ., London, 1993)
 \bibitem{wolfram} Weisstein, Eric W., A Wolfram Web Resource,\url{http://functions.wolfram.com/ElementaryFunctions/Log/06/01/05/01/}
 
 
 
 \end{thebibliography}
\end{document}